\documentclass[aps,prl,superscriptaddress,twocolumn,reprint,amsmath,amssymb]{revtex4-2}

\usepackage[english]{babel}
\usepackage{graphicx}
\usepackage[dvipsnames]{xcolor}
\usepackage[colorlinks=true,urlcolor=blue,linkcolor=blue,citecolor=blue]{hyperref}
\usepackage{epstopdf}
\usepackage{dcolumn}
\usepackage{bm}
\usepackage{epsfig}
\usepackage{soul}
\usepackage{multirow}
\usepackage[caption=false]{subfig}
\usepackage{wrapfig}

\begin{document}

\title{Nuclear magnetic resonance spectroscopy in pulsed magnetic fields}

\author{H.~K\"{u}hne}
\affiliation{Helmholtz-Zentrum Dresden-Rossendorf, Hochfeld-Magnetlabor Dresden (HLD-EMFL), 01328 Dresden, Germany}

\author{Y. Ihara}
\affiliation{Department of Physics, Faculty of Science, Hokkaido University, Sapporo 060-0810, Japan}

\date{\today}

\begin{abstract}
This article provides an introduction to nuclear magnetic resonance spectroscopy in pulsed magnetic fields (PFNMR), focusing on its capabilities, applications, and future developments in research involving high magnetic fields. It highlights the significance of PFNMR in enhancing the understanding of solid-state materials, with particular emphasis on those exhibiting complex interactions and strong electronic correlations. Several technical aspects are discussed, including the challenges associated with high-frequency NMR experiments. The power of PFNMR is showcased through several examples, including studies on the topical materials LiCuVO$_4$, SrCu$_2$(BO$_3$)$_2$, and CeIn$_3$, offering insights into their magnetic and electronic properties at high magnetic fields. The article also discusses possible future directions for the technique, including improvements in PFNMR instrumentation and the exploration of materials under extreme conditions. This exposition underscores the role of PFNMR in advancing the frontiers of materials-science research.
\end{abstract}
\maketitle

\section{1. Introduction}

Since the pioneering experiments performed in the mid-1940s by Purcell, Torrey, and
Pound in Cambridge, and by Bloch, Hansen, and Packard in Stanford \cite{Bloch1946,Purcell1946}, nuclear magnetic resonance (NMR) spectroscopy has undergone continuous advancements and developments. Today, it is well established as a versatile and widely applicable spectroscopic method  across various scientific disciplines.

In fundamental condensed-matter research, NMR spectroscopy is a powerful tool for exploring the electronic and magnetic properties of materials. It facilitates the accurate measurement of nuclear spin-lattice relaxation times, nuclear spin-spin interactions, and electronic spin densities, thus advancing the understanding of condensed-matter phenomena such as superconductivity, magnetism, and phase transitions.
In chemistry, NMR spectroscopy is indispensable for structural determination and chemical analysis. It provides comprehensive insights into molecular environments, bond connectivity, and stereochemistry. This enables the identification and detailed characterization of both organic and inorganic compounds, as well as complex biomolecules such as proteins, nucleic acids, and carbohydrates.
Moreover, NMR spectroscopy is instrumental in biological and medical research, providing non-invasive methods to study biological systems. In medical diagnostics, magnetic resonance imaging (MRI) utilizes NMR principles to create detailed anatomical images of tissues and organs. This aids in the detection and diagnosis of various diseases, from cancer to neurological disorders.

NMR spectroscopy also has broad industrial applications due to its ability to analyze the composition and properties of materials. In metallurgy, it is used to assess metal alloys, supporting quality control and enhancing manufacturing processes. In the food industry, NMR techniques verify the authenticity and purity of food products, identify adulteration, and track chemical changes during processing and storage. Additionally, NMR is vital in environmental science and sustainability, aiding in the characterization and remediation of pollutants, as well as in monitoring soil and water quality and developing effective recycling methods for waste materials.

NMR operates through the resonant excitation of nuclear spin ensembles in a sample, whether solid, liquid, or gaseous, followed by the detection of their response or relaxation back to thermal equilibrium. The NMR signal provides extensive microscopic data about the sample. For liquids, this includes details about chemical composition and molecular structure. For solids, it offers, for example, insight into the microscopic arrangements and interactions of magnetic moments. To achieve this, the nuclear moments of the chemical isotopes inherent to the material act as microscopic detectors, probing the local magnetic fields and gradients of crystal electric fields caused by the electronic degrees of freedom.

\begin{figure*}[!ht]
	\centering
	\includegraphics[width=0.7 \linewidth]{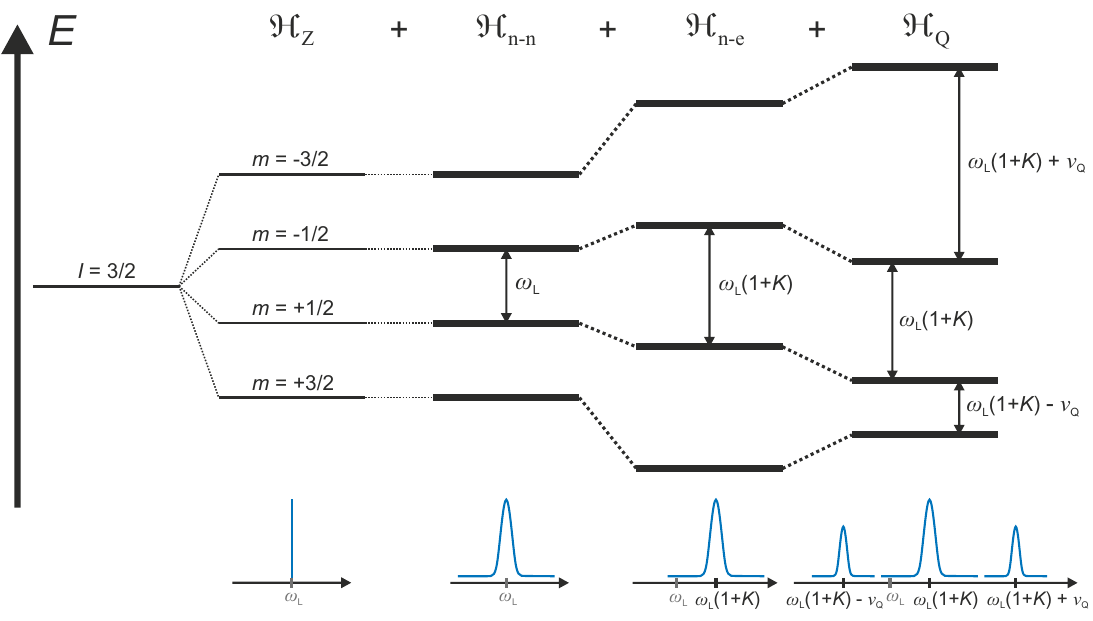}
	\caption{\label{Energy_diagram} Summary of all internuclear and nuclear-electron interactions, their influence on the nuclear energy levels and the resulting NMR spectra for a nuclear isotope with angular momentum $I = 3/2$.}
\end{figure*}

While the concepts presented in this article are general in nature and apply to various applications in NMR experiments, in the following the primary focus will be on investigations of solid-state materials in condensed-matter physics, as these are the cases where NMR spectroscopy in pulsed magnetic fields is currently most commonly used.

\subsection{NMR in a nutshell}

Several excellent textbooks provide comprehensive introductions to the concepts and applications of NMR \cite{Abragam1961,Fukushima1981,Mehring1983,Slichter1990}. In the following, we introduce some fundamental concepts of NMR that are essential for understanding PFNMR. The Hamiltonian that describes the internuclear and nuclear-electron interactions can be written as:
\begin{equation}
\mathcal{H} = \mathcal{H}_{z} + \mathcal{H}_{n-n} + \mathcal{H}_{n-e} + \mathcal{H}_{Q},
\end{equation}
where $\mathcal{H}_{z}$ represents the Zeeman interaction between a nuclear moment and an external magnetic field, $\mathcal{H}_{n-n}$ denotes the interaction between nuclear moments, $\mathcal{H}_{n-e}$ describes the interaction between the nuclear moments and the spins and orbital moments of the electrons, and $\mathcal{H}_{Q}$ accounts for the interaction between the quadrupole moment of the nucleus and the local gradient of the crystal electric field. The latter is relevant only for nuclei with an angular momentum $I > 1/2$, situated in a non-cubic environment. Since it is also common in the literature to refer to $I$ as the nuclear spin, both terms are used synonymously in this article.

Figure~\ref{Energy_diagram} illustrates the nuclear energy-level diagram and the corresponding NMR spectra for a nuclear isotope with $I = 3/2$, with some specific interaction parameters omitted for brevity. Most NMR setups involve a laboratory magnet that generates a strong magnetic field $\mu_0 H_0$ of several tesla. This field induces a nuclear Zeeman splitting, resulting in the polarization of the nuclear spin ensemble. The interaction of the nuclear spin with the applied magnetic field lifts the degeneracy among the $2I + 1$ nuclear energy levels. Correspondingly, a delta peak appears in the NMR spectrum at the Larmor frequency $\omega_L$, indicating a transition between two adjacent energy levels. Interactions among the various nuclear moments, denoted as $\mathcal{H}_{n-n}$, broaden these energy levels, resulting in a Gaussian-shaped resonance line, centered at $\omega_L$. 

Magnetic hyperfine interactions between the nuclear moments and the spins and orbital moments of electrons cause a frequency shift of the resonance line, an effect known as the Knight shift $K$ in condensed-matter physics. Furthermore, electric-quadrupole interactions induce an additional shift in the energy levels with a characteristic frequency $\nu_Q$, effectively removing the degeneracy of the transitions. This results in the splitting of the NMR spectrum into three distinct lines, with the central transition remaining unshifted in first-order approximation.

The distribution of nuclear states in a statistical ensemble at thermodynamic equilibrium conforms to the Boltzmann distribution, expressed as $P(m) = e^{-E_m/k_BT}$, where $E_m$ represents the nuclear energy levels. The disparity in population between these energy levels results in a nuclear spin polarization, which corresponds to a net nuclear magnetization, denoted as \textbf{M}, oriented along the direction of the magnetic field $\textbf{H}_{0}$.

\begin{figure}[tb]
	\centering
	\includegraphics[width=0.81\linewidth]{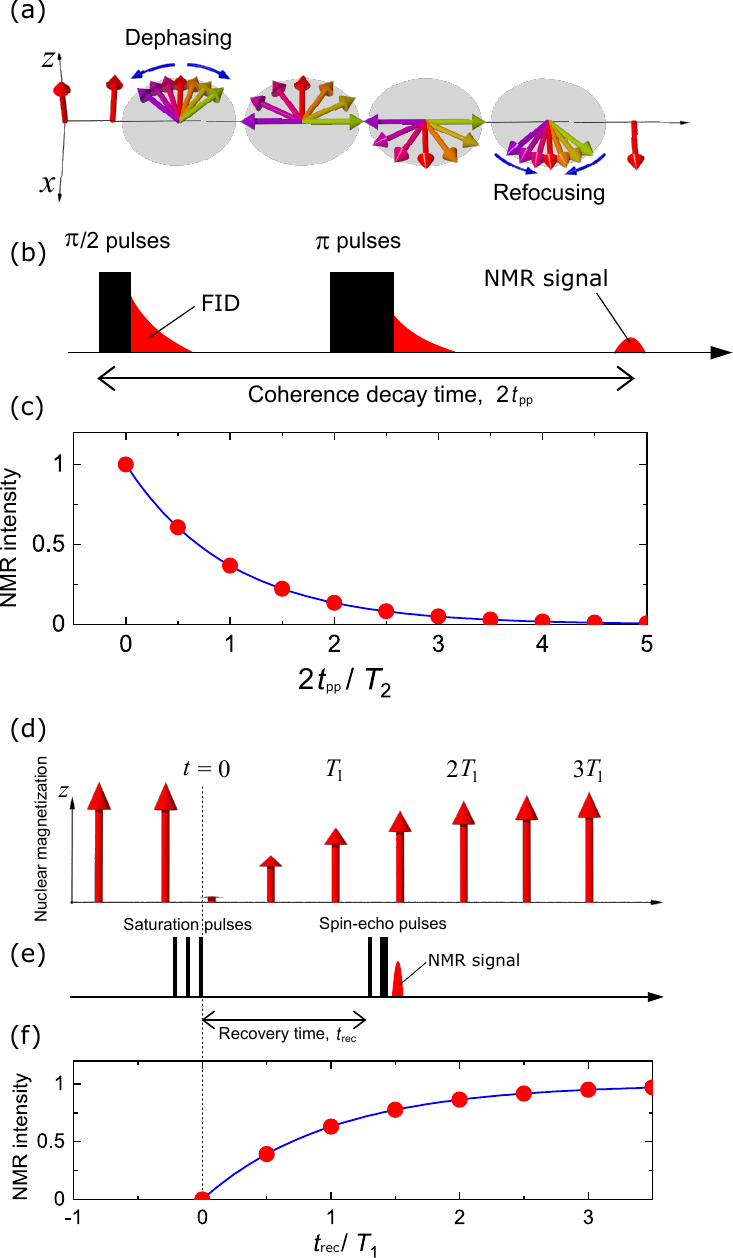}
	\caption{\label{T2T1Meas} Standard methods for measuring the NMR relaxation times $T_2$ $\left[ (a)-(c) \right]$ and $T_1$ $\left[ (d)-(f) \right]$. (a) Time evolution of the nuclear spin ensemble, represented by colored arrows, and (b) corresponding spin-echo pulse sequence. The spin directions, dephasing after a $\pi/2$ pulse, are refocused by a $\pi$ pulse at $t = t_{pp}$, and a spin-echo signal is detected at $t = 2 t_{pp}$. (c) Decay of the NMR signal intensity with a characteristic timescale $T_2$, recorded by repeating the sequence in (b) with various delays $t_{pp}$. (d) Red arrows indicate the longitudinal nuclear magnetization at specific times. Immediately after the saturation pulses at $t=0$, the nuclear magnetization is reduced to zero and then gradually recovers over time. (e) Saturation-recovery pulse sequence for measuring $T_1$. The recovering nuclear magnetization is determined from the NMR signal intensity, which is probed by the spin-echo pulse group after the recovery time, $t_{rec}$. (f) Recovery of the signal intensity with a characteristic timescale $T_1$, obtained by repeating the sequence in (e) with varying delay $t_{rec}$.}
\end{figure}

By applying a time-varying radio-frequency (RF) field $\textbf{H}_{RF}$, orthogonal to $\textbf{H}_{0}$, transitions with $\Delta m = ±1$ between the nuclear eigenstates are induced. The resonance condition for these transitions is given by the equation:
\begin{equation}
\Delta E = \hbar \omega_L = \hbar \cdot \gamma_n H_0,
\end{equation}
where $\Delta E$ denotes the energy difference between the eigenstates, $\omega_L = \gamma_n H_0$ is the Larmor frequency of the nuclear moments, and $\gamma_n$ is the isotope-specific nuclear gyromagnetic ratio.
In most NMR experiments, the nuclear Larmor frequency is in the megahertz (MHz) to the gigahertz (GHz) regime, depending on the isotope, the strength of the applied magnetic field, and the electronic characteristics of the sample under examination.

In order to discuss the dynamics of the nuclear spin system, we consider the simple case of nuclear moments characterized by $I = 1/2$. In this case, the behavior of the nuclear magnetization can be described using a classical framework. 
The nuclear magnetic moment, denoted as $\bm{\mu} = \gamma_n \hbar \textbf{I}$, when placed in an external magnetic field $\textbf{H}_{0}$, experiences a torque $\bm{\mu} \times \textbf{H}_{0}$. Thus, the classical equation governing the motion of the nuclear magnetization $\textbf{M}$, which is the sum of individual nuclear magnetic moments $\bm{\mu}_i$, can be expressed as
\begin{eqnarray}
\frac{d\mathbf{M}}{dt}=\gamma_n (\mathbf{M}\times \mathbf{H_0})~.
\end{eqnarray}
The solution of this equation describes the precession of the magnetization $\textbf{M}$ about the external magnetic field $\mathbf{H_0}$ at a frequency $\omega_L = \gamma_n \mathbf{H_0}$. 
Within the laboratory coordinate system, the $z$ axis is aligned parallel to the magnetic field. The components of the magnetization $\mathbf{M}$ achieve thermodynamic equilibrium as $M_z = M_0$ and $M_x = M_y = 0$. Deviations from this equilibrium state, induced by $\mathbf{H}_{RF}$, lead to relaxation dynamics characterized by two distinct timescales: the longitudinal and transverse relaxation times, $T_1$ and $T_2$, respectively. The dynamics of nuclear magnetic moments under the influence of a local magnetic field $\mathbf{H}$, comprising both static and radio-frequency components $\mathbf{H}_0$ and $\mathbf{H}_{RF}$, can be conceptualized as free spin motion and relaxation-driven motion, as described by the phenomenological Bloch equations:
\begin{eqnarray}
\frac{dM_x}{dt}&=&\gamma_n (\mathbf{M}\times \mathbf{H})_x-\frac{M_x}{T_2}\\
\frac{dM_y}{dt}&=&\gamma_n (\mathbf{M}\times \mathbf{H})_y-\frac{M_y}{T_2}\\
\frac{dM_z}{dt}&=&\gamma_n (\mathbf{M}\times \mathbf{H})_z - \frac{M_z - M_0}{T_1}
\end{eqnarray}
In the case of an individual nuclear spin, both classical and quantum-mechanical frameworks yield identical outcomes. For a quantum-mechanical description of a statistical ensemble of spins, it is necessary to introduce a statistical operator $\rho$, which is represented by a density matrix.

The mechanisms determining ${T_1}$ and ${T_2}$ are given by the physical properties of the material under examination. In the context of solid-state samples exhibiting electronic correlations, ${T_1}$ is often conceptualized as the characteristic thermalization time between the excited thermodynamic ensemble of the nuclear spin system and the low-energy excitations of the electronic system. As such, ${T_1}$ represents a measure of, for example, the correlation length of localized magnetic moments in low-dimensional spin systems, or the density of low-energy quasiparticles in metals.
On the other hand, ${T_2}$, the transverse coherence time, is determined by various mechanisms. These include, but are not confined to, interactions between nuclear spins or diffusion processes within the material.

In a conventional steady-field NMR experiment, the nuclear spin-spin relaxation time ${T_2}$ is measured by recording the intensity of the nuclear spin echo for various pulse–pulse separation times, $t_{pp}$, as illustrated in Figs.~\ref{T2T1Meas} (a)--\ref{T2T1Meas}(c). The spin-echo method retrieves the NMR signal through the procedure depicted in Fig.~\ref{T2T1Meas}(a). Initially, at $t = 0$, a $\pi/2$ pulse rotates the nuclear spins 90$^\circ$ away from the external field direction. Subsequently, the spins begin to dephase in the $xy$ plane due to variations of the Larmor frequency among the spins, which are attributed to the experimentally unavoidable distribution of static fields throughout the sample volume. During this period ($0 < t < t_{pp}$), the nuclear
spins precess in the $xy$ plane, generating a free induction decay (FID) signal. Given that the primary axis of the RF coil, which contains the sample, is oriented perpendicular to the external magnetic field, only the $xy$ components of the nuclear magnetization induce a voltage signal. The FID signal decays on a timescale $T^*_2$, determined by the static-field distribution, which also determines the linewidth of the NMR spectrum. Although the FID signal intensity drops to very small values for $t > T^*_2$, the NMR signal can be refocused with a $\pi$ pulse, provided that the phase coherence of the nuclear spin ensemble is preserved. The application of the
$\pi$ pulse at $t = t_{pp}$ rotates the nuclear spins by 180$^\circ$ about the $y$ axis, causing the dephased spins to return to the original direction, following their trajectories prior to the $\pi$ pulse.
For the spins to be refocused, the magnetic-field distribution leading to the decay in the FID signal must remain unchanged throughout the coherence decay time of $2 t_{pp}$. If this condition is not met, the diffused nuclear spins do not revert to their initial orientation, resulting in a loss of spin-echo intensity. The spin-echo decay curve of the detected signal intensity, $m_n(t)$, can be described by an exponential function $m_n(t) = m_n(0) e^{- 2 t_{pp}/T_2}$ in the case of weak spin–spin interactions, as shown in Fig.~\ref{T2T1Meas}(c). Conversely, for strong spin–spin interactions, the decay curve can be described by a Gaussian-type form.

In order to measure the nuclear spin-lattice relaxation time ${T_1}$ using the saturation-recovery method, the pulse sequence shown in Fig.~\ref{T2T1Meas}(e) can be employed. Initially, at $t = 0$, one or several saturation pulses are applied to the sample, effectively reducing the $z$ component of the nuclear magnetization to zero. Therefore, no signal is detected immediately after the application of the saturation pulses. Over time, the nuclear magnetization gradually returns to its thermal equilibrium value, $m_n(\infty)$, on a timescale characterized by the relaxation time $T_1$. After a delay $t_{rec}$, a spin-echo pulse group is applied, generating an echo NMR signal. The magnitude of the recovered nuclear magnetization is inferred from the intensity of this NMR signal, which is directly proportional to the longitudinal component of the nuclear magnetization before the application of the spin-echo pulse group.
By varying the recovery time $t_{rec}$, this process yields $T_1$. Fig.~\ref{T2T1Meas}(f) illustrates this for the case of a nuclear spin $I = 1/2$, where the recovery curve can be described by a single-exponential function 
$m_n(t) = m_n(\infty) \left[ 1 - e^{- t_{rec}/T_1} \right]$.

\begin{figure}[tbp]
	\centering
	\includegraphics[width=0.99\linewidth]{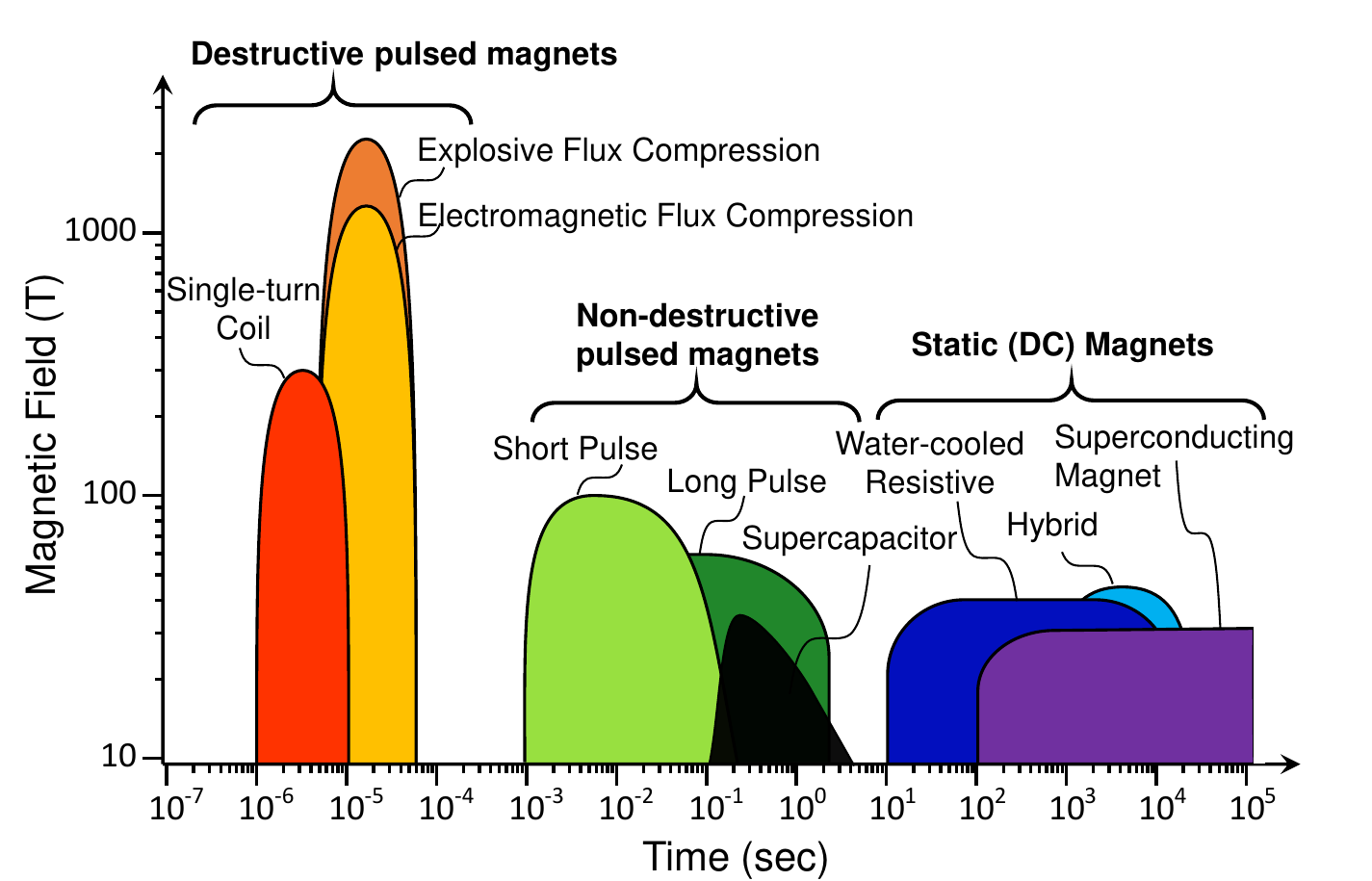}
	\caption{\label{FieldComparison} Comparison of magnetic-field generation techniques in terms of achievable field strength and duration (see Ref. \cite{Matsui2021} and citations therein).}
\end{figure}

\subsection{NMR at high magnetic fields}

In most cases, NMR measurements greatly benefit from the use of high magnetic fields. This is because both the spectral resolution and the signal intensity are enhanced by a large Zeeman splitting and the resulting population imbalance of the nuclear energy levels, respectively. 
In most NMR experiments, the laboratory magnetic fields are generated by superconducting solenoid magnets. These magnets provide field characteristics that are exceptionally stable over time and exhibit very high spatial homogeneity throughout the sample volume. The specific field requirements depend on the linewidth of the NMR spectrum and the characteristic nuclear relaxation times of the material under study. For instance, for NMR studies on liquids, resolving the weak internal fields may require a relative field variation on the order of $\Delta H / H_0 \approx 10^{-9}$ across the sample volume. Conversely, solid-state materials with internal magnetism can exhibit intrinsic NMR linewidths on the order of $\Delta H / H_0 \approx 10^{-3}$. The latter case significantly reduces the demands on the precision and, consequently, the costs of magnet production. Similarly, solid-state samples with short relaxation times may not require magnets that maintain a high temporal stability of the magnetic field.

Figure \ref{FieldComparison} shows an overview of the various technological approaches for generating highest magnetic fields in condensed-matter research.
Currently, commercially available NMR spectrometers using superconducting magnets can achieve fields up to approximately 28 tesla. Recent technological advancements have enabled experiments with all-superconducting magnets incorporating high-temperature superconductors in specialized user facilities, achieving somewhat higher fields. For fields beyond these values, NMR experiments can be performed using resistive magnets or hybrids of resistive and superconducting magnets, reaching up to about 45 tesla. 
However, the operational costs for such setups are high, and there are currently only very few locations worldwide, where experiments can be conducted at these maximum static fields. 

Using destructive pulsed-field magnets, which fragment radially away from the sample during the magnetic-field pulse, it is possible to achieve fields that reach several hundred tesla. 
While the cryostat and the sample remain undamaged after the coil fragmentation in the single-turn coil method, the electromagnetic and explosive flux compression methods can result in the destruction of several parts of the experimental setup, including the sample \cite{Miura2003}.
Employing these techniques to measure the physical properties of materials poses numerous challenges, primarily due to the extremely short pulse durations, often lasting only a few microseconds, rendering these magnets unsuitable for pulsed-field NMR experiments.

Looking ahead, performing NMR at magnetic-field strengths surpassing those of steady-field experiments necessitates the use of nondestructive pulsed magnets, capable of reaching fields up to about 100 tesla. In such experiments, NMR measurements are performed during a brief magnetic-field pulse, typically lasting several tens of milliseconds. Due to the substantial installation costs and stringent safety requirements associated with the required energies, facilities with the strongest pulsed magnets are located in only a few specialized high-field laboratories worldwide. The technology of these magnets is subject to ongoing advancements. Moreover, the range of experimental techniques that can be conducted in these magnets has significantly broadened and matured over recent decades.

\begin{figure}[tbp]
	\centering
	\includegraphics[width=0.95\linewidth]{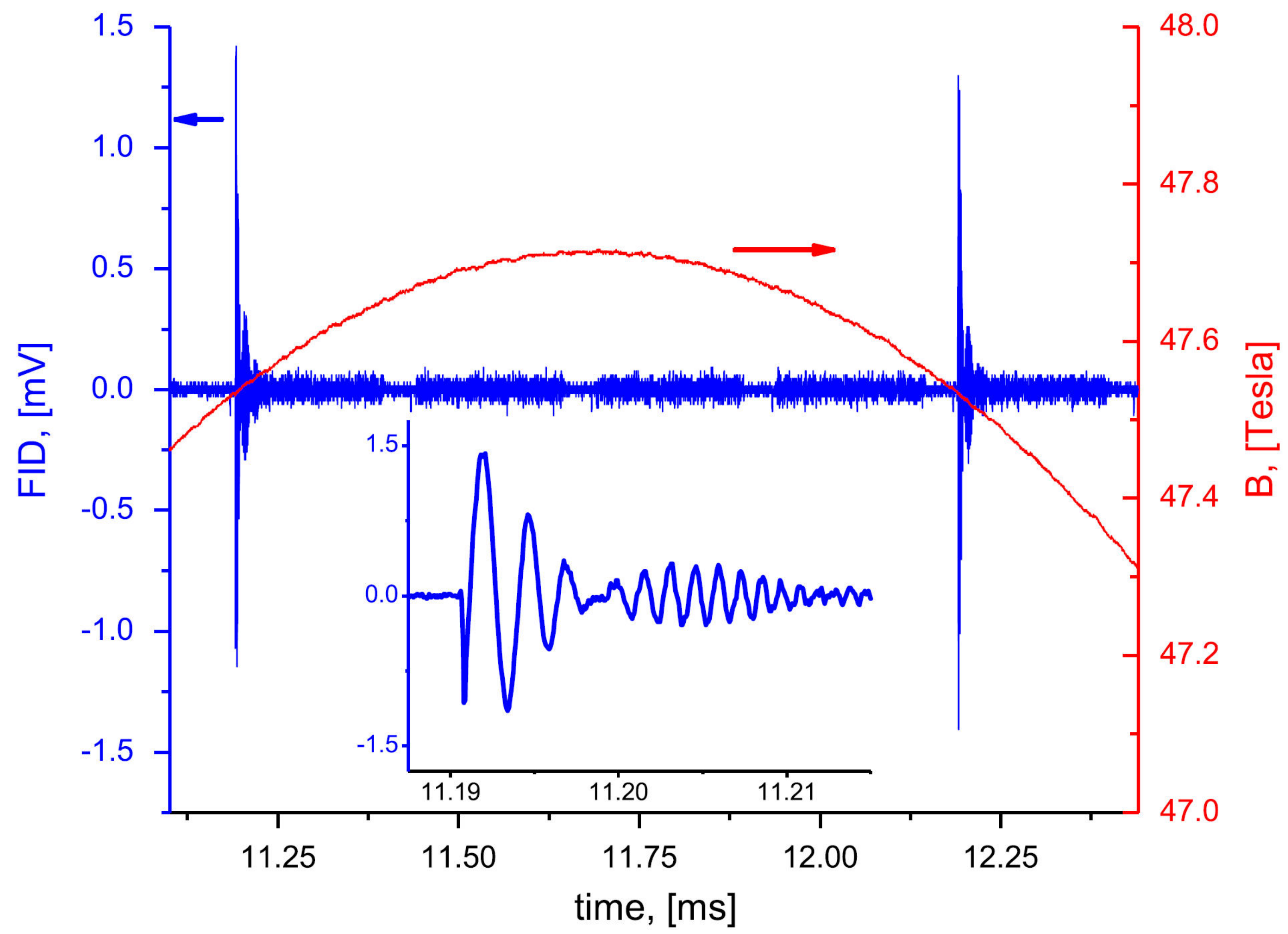}
	\caption{\label{Haase2GHzFigure} $^1$H NMR signal at 2.027 GHz within a 50 T pulsed-field magnet. The red curve shows the pulsed field near its peak, whereas the blue curve displays a five-segment NMR receiver trace (real part) over the same timeframe; an FID signal is observed following the first and fifth RF pulses. Inset: an enlarged view of the first FID signal. Reproduced with permission from Haase \textit{et al.}, Solid State Nucl. Magn. Reson. \textbf{27}, 206 (2005). Copyright 2005 Elsevier \cite{Haase2005}.}
\end{figure}

Pioneering the development of NMR in pulsed-field magnets, the first experiments were performed by J. Haase et al. in the early 2000s \cite{Haase2003,Haase2004,Kozlov2005,Haase2005}. These initial studies were subsequently expanded upon in laboratories in Germany, France, Japan, and China \cite{Zheng2009,Hamad2011,Meier2011,Meier2012,Stork2013,Kohlrautz2016,Orlova2016,Orlova2017,Kohlrautz2016SCBO,Tokunaga2019,Ihara2021,Matsui2021,Kohama2022,Wei2023}. An example of the first reported $^1$H FID signals at approximately 2 GHz is shown in Fig.~\ref{Haase2GHzFigure}. Beyond $^1$H NMR experiments at very high frequencies, data were also recorded on solid samples, demonstrating the feasibility of measuring relaxation times as well as precise frequency shifts utilizing internal NMR references. Subsequent studies by various research groups facilitated the examination of internal-field distributions within complex materials, not only at high fields but also at cryogenic temperatures. Some of these developments will be discussed in Section 3 of this article.

The number of applications utilizing PFNMR is expected to increase due to continuous advancements in instrumentation and a growing body of scientific studies. Over time, PFNMR may be established as a standard technique for adressing topical and unresolved issues in solid-state physics, particularly for probing electronic ground states stabilized in high magnetic fields, by providing microscopic insights into the interactions and polarization of electronic moments. Furthermore, PFNMR may provide improved spectral resolution or allow for differentiation between paramagnetic effects and those independent of magnetic fields, such as effects arising from local electric-field gradients.

The information from PFNMR is complementary to that from thermodynamic methods, such as measurements of the magnetization or the magnetocaloric effect, which are also applicable to investigate materials in pulsed magnetic fields. Although there are reports on first experimental studies that combine pulsed magnetic fields with x-rays and neutrons, PFNMR is anticipated to remain, for the foreseeable future, one of the few methods capable of obtaining microscopic information at the highest magnetic fields, and it is likely the sole technique for fields approaching the regime of 100 tesla.

\section{2. Experimental procedures}

\subsection{Pulsed-field generation}

The primary method for generating strong magnetic fields with pulsed-field magnets involves directing a large current through a solenoid coil. These currents are typically generated by rapidly discharging electrical energy from a large capacitor bank into a low-impedance magnet coil. The currents employed are substantially higher than those for superconducting magnets, which are limited to a few hundred amperes to avoid destabilization of the superconducting state due to critical current thresholds.

\begin{figure}[tbp]
	\centering
	\includegraphics[width=0.95\linewidth]{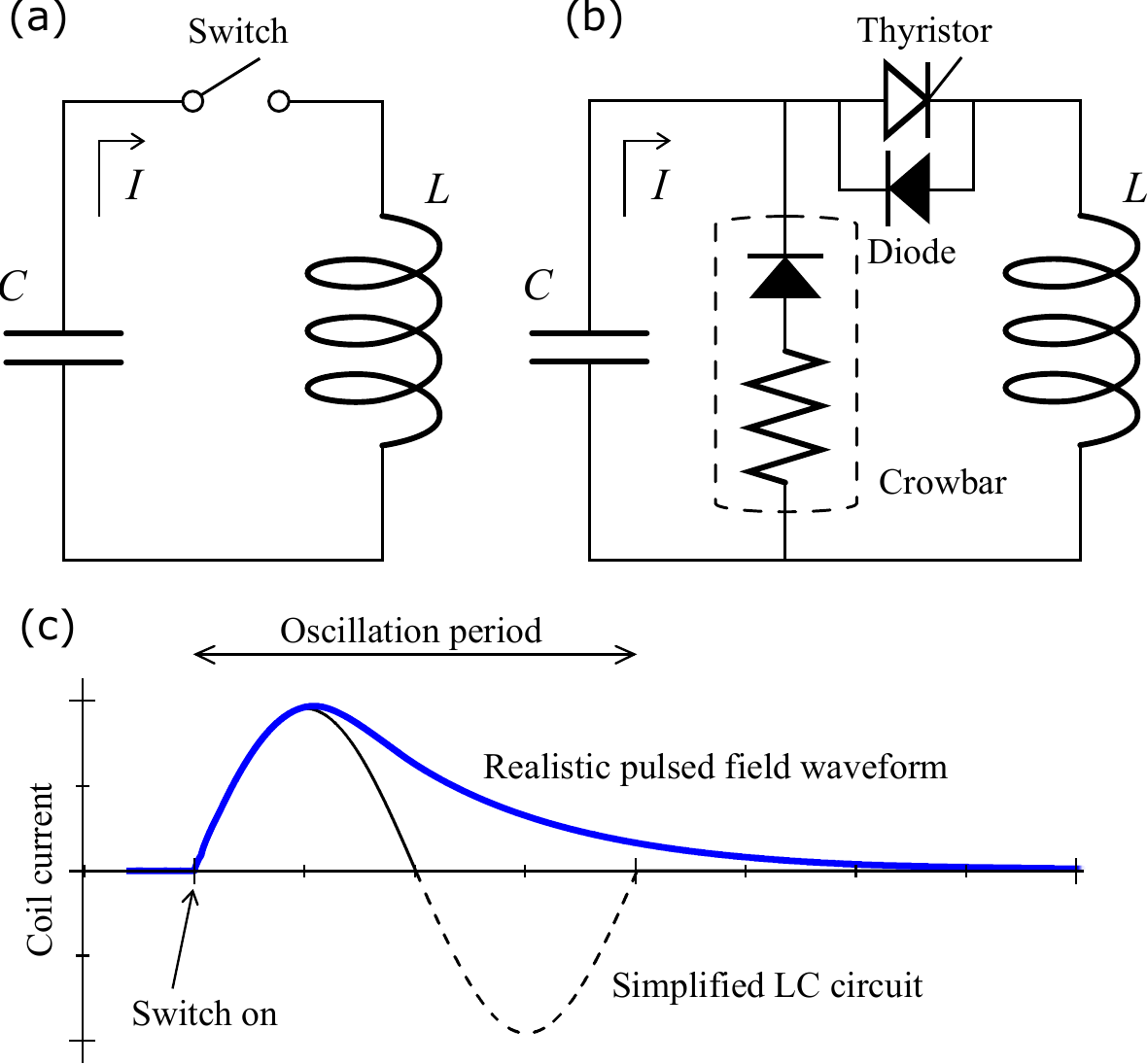}
	\caption{\label{LCcircuit} (a) An electric $LC$ circuit. Initially, electric charge is stored in a capacitor $C$. Upon closing the switch, a current starts to oscillate between the capacitor and the inductor $L$ with a period of $2\pi\sqrt{LC}$. (b) Simplified scheme of a pulsed-field installation. Here, a thyristor functions as a high-speed switch, capable of handling large currents and high voltages. A crowbar circuit, incorporating a crowbar diode and a pulse-shaping resistor, ensures an aperiodic current decay, thereby preventing substantial current backflow into the capacitors. (c) Time dependence of the coil current. The black solid line represents the current profile for the simple $LC$ circuit in (a). In contrast, the blue line illustrates a realistic pulsed-field profile, corresponding to the approximated pulsed-field circuit described in (b).}
\end{figure}

As illustrated in Fig.~\ref{LCcircuit}(a), a basic $LC$ circuit consists of a capacitor, which stores electric charge, and an inductor, which generates a magnetic field upon current flow. The periodicity of the oscillating current in this circuit is determined by the system's capacitance ($C$) and inductance ($L$), and is given by $2 \pi\sqrt{LC}$. A slightly more complex circuit, representing a simplified approximation of an infrastructure for pulsed-field experiments, is shown in Fig.~\ref{LCcircuit}(b).
In this scheme, a thyristor functions as a high-speed switch, capable of handling large currents and high voltages. Additionally, a crowbar diode and a pulse-shaping resistor are placed in parallel to the capacitor modules and the magnet coil.

As depicted by the blue line in Fig.~\ref{LCcircuit}(c), the coil current begins to flow when the thyristor switch is activated, allowing current to flow from the capacitor bank to the pulsed-field magnet. During the rising part of the coil current, a current flow through the crowbar circuit is blocked by the diode. After the peak of the coil current and pulsed magnetic field, respectively, the voltage across the crowbar circuit is reversed. Consequently, the current reversal is controlled via the crowbar circuit, shaping the current decay close to the aperiodic limit in order to avoid a substantial, harmful backflow into the capacitor modules.

Several factors limit the maximum achievable magnetic field, including Joule heating, electromagnetic forces, and the maximum energy stored in the capacitor.
Joule heating of the magnet, generated by the currents of several thousand amps, is substantial even if the resistance of the magnet wire is as low as a few milliohms, since it is proportional to $RI^2$. 
During the brief duration of pulsed-field generation, the finite thermal resistance prevents heat dissipation to the thermal bath.
As a result, the nitrogen-cooled magnet coil, initially at 77 K, may warm up close to room temperature after a single field pulse.

\begin{figure}[tbp]
	\centering
	\includegraphics[width=0.95\linewidth]{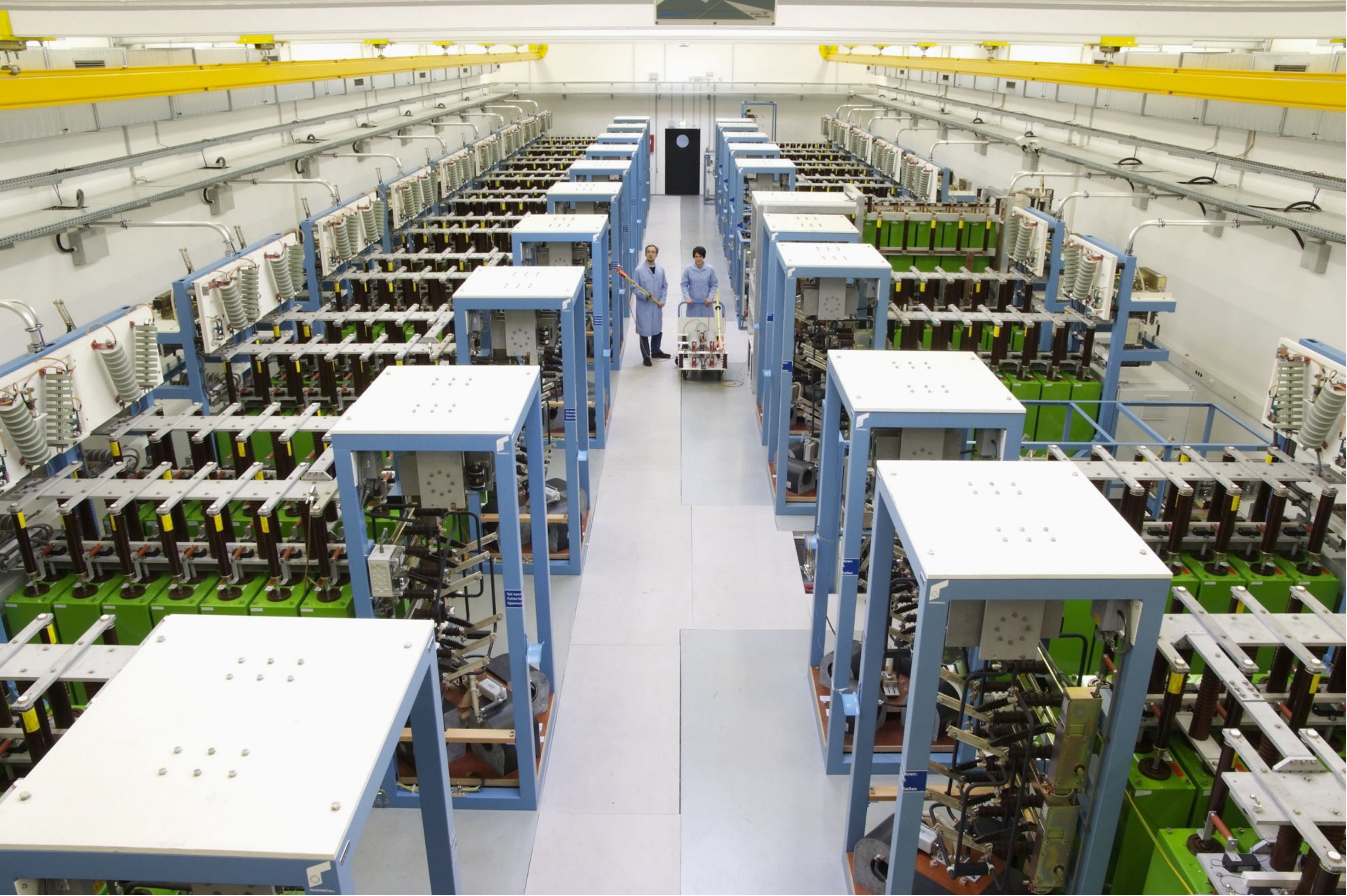}
	\caption{\label{Cap_bank_HLD} Photograph of the 49 MJ capacitor bank at the Dresden High Magnetic Field Laboratory. The hall contains 20 capacitor modules, which can be interconnected flexibly depending on the requirements of the pulsed-field experiment.}
\end{figure}

The electromagnetic forces that cause radial expansion of the coil are considerable. 
The tensile strength of pure copper is exceeded by the force exerted on the magnet wire at around 25 T, which can lead to deformation or, in severe cases, complete failure of the magnet. To mitigate this, reinforced wire and multiple support layers are incorporated in the construction of a pulsed-field magnet. The capacitors used in pulsed-field experiments can withstand high voltages of up to several tens of kilovolts but have a relatively low energy density. Consequently, large facilities, such as the Dresden High Magnetic Field Laboratory (HLD), require extensive capacitor arrays housed in sizable halls, as shown in Fig.~\ref{Cap_bank_HLD}.

A key parameter in PFNMR experiments is the duration of the pulsed magnetic field. 
The pulse duration, which is of the order of $\pi\sqrt{LC}$, can be increased by either raising the inductance or the capacitance.
For example, at the HLD, large capacitances allow for pulse durations that exceed 100 ms. This is advantageous for PFNMR, as it minimizes temporal field variations during the NMR experiment, which typically lasts about 0.1 ms. 
For instance, to detect a $^{63}$Cu NMR signal ($\gamma=11.285$ MHz/T) with a receiver frequency window of 1 MHz, the field variation within the time of the NMR pulse sequence of 0.1 ms must be less than 0.088 T. 
The approximately sinusoidal field profile at the field maximum facilitates data recording with minimal temporal changes in the magnetic field. Consequently, most PFNMR experiments are performed in this peak regime. 

\begin{figure}[tbp]
	\centering
	\includegraphics[width=0.75\linewidth]{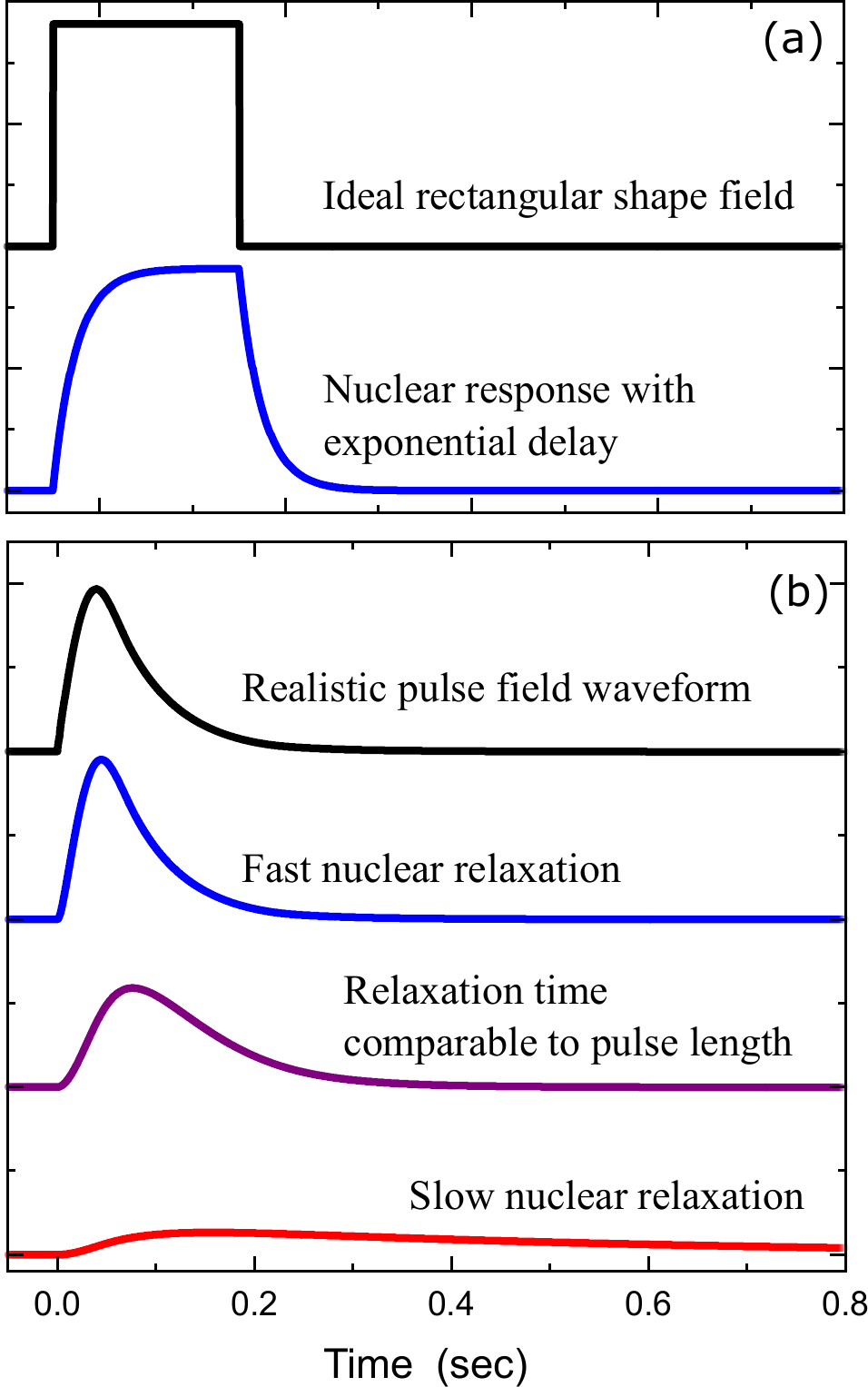}
	\caption{\label{delayedresponse} Temporal evolution of nuclear magnetization in response to various pulsed magnetic fields. (a) When the external field is applied as an idealized step function, the nuclear magnetization responds with an exponential relaxation. (b) In a more realistic scenario, where the $T_1$ relaxation time is sufficiently short (top, in blue), the nuclear magnetization closely follows the waveform of the pulsed field (shown in black). If the relaxation time is comparable to the time scale of the pulsed field (middle, in purple), the nuclear magnetization cannot attain thermal equilibrium. Conversely, if $T_1$ is very long in comparison (bottom, in red), only a small fraction of the fully polarized nuclear magnetization is achieved.}
\end{figure}

Long-duration field pulses are essential for studying materials with long nuclear relaxation times. As $T_1$ represents the timescale in which the nuclear magnetization reaches thermal equilibrium, the nuclear magnetic moments follow the pulsed external field with a delayed response characterized by $T_1$. Thus, if the duration of a pulsed magnetic field is significantly shorter than $T_1$, the nuclear magnetization does not reach a sufficiently polarized state to conduct a successful PFNMR experiment, as illustrated in Fig.~\ref{delayedresponse}.

The response of the nuclear magnetization is slow if the nuclear moments are only weakly coupled to the electronic system or if the spectral density of the low-energy electronic moment fluctuations is small. 
In this context, PFNMR is suited to investigate, for example, the field-induced magnetic structures in materials with relatively short $T_1$, rather than achieving high spectral resolution or detecting sharp spectral lines in materials with long $T_1$.  
When $T_1$ is comparable to the pulse duration, the resulting NMR signal is weaker than what would be observed in a fully polarized state under a steady magnetic field. However, even under these conditions, the relative spectral line separation and intensities recorded near the peak of a pulsed field can be used to obtain broadband field-sweep NMR spectra, particularly if the nuclear magnetization is in a quasi-stationary state, as will be discussed further below. 

Given the slow variation of the external magnetic field and nearly constant nuclear magnetization at the peak of the pulsed magnetic field, NMR measurements are typically feasible only near this maximum. However, this is not a drawback since measurements at the highest possible field are a primary goal of PFNMR experiments. 
Detecting the NMR signal is challenging because the resonance condition is met only within a very narrow field range with respect to the small RF bandwidth around a given center frequency. 
Even in magnetic materials, where the NMR spectra are generally broad, the relative field range satisfying resonance conditions is usually less than 1 \%. This requires the peak of the pulsed magnetic field to be reproducibly controlled within less than 1 \% variation. Ideally, the peak field should slightly exceed the resonant conditions to ensure that the NMR signal is detected. Conversely, if the field does not reach the necessary strength, a resonance will not be observed.
Achieving such precise control over the peak field is challenging due to two main factors.
The first is achieving accurate voltage control when charging the capacitors up to the kilovolt range. Fortunately, modern power supplies can regulate high voltages with precision much better than 1 \%. 
The second challenge involves managing the temperature of the pulsed-field magnet. 
The magnet coil is cooled with liquid nitrogen prior to a pulse to lower the resistance of the wire. However, the temperature of the magnet does not equal the nitrogen boiling temperature when the cooldown is not complete. During a pulse, the temperature of the coil wire increases significantly due to Joule heating. 
This requires the magnet to be cooled back to initial conditions before another pulse can be generated. The temperature-dependent resistance of the wire may lead to small variations in the initial resistance for subsequent field pulses, affecting the peak fields achieved. Thus, monitoring the coil resistance just before a pulse is crucial for consistent peak-field generation.
Through careful calibration of the magnetic field for each combination of capacitors and pulsed-field magnet, reliable reproducibility of the peak magnetic field can be achieved, facilitating the acquisition of high-fidelity NMR spectra. 

\begin{figure}[tbp]
	\centering
	\includegraphics[width=0.95\linewidth]{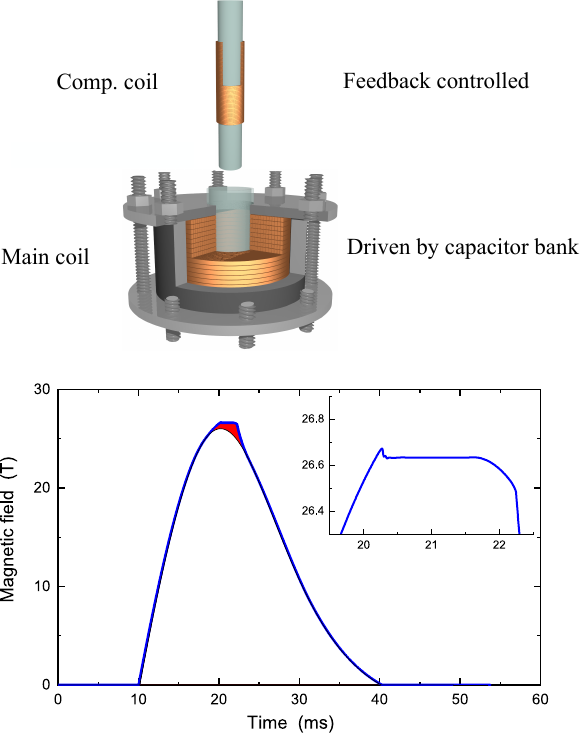}
	\caption{\label{flattop} Top: Schematic drawing of a compensation coil and main coil. The small compensation coil is positioned within the larger main coil, producing dynamically controlled magnetic fields of up to approximately 2 T. Bottom: Profile of a flat-top pulse field. The field generated by the compensation coil is depicted in red. In addition to the large magnetic field from the main coil, an additional smaller field from the compensation coil dynamically adjusts the overall pulsed-field profile, achieving a flat-top regime with a relative field variation of about 100 ppm. The controlled, flat region near the peak is magnified in the inset for clarity.}
\end{figure}

Another method for achieving reproducible, well-controlled field profiles is the flat-top field approach, where the field strength is dynamically controlled during the pulse \cite{Kohama2015}, as shown in Fig.~\ref{flattop}. 
Originally developed to create a quasi-steady field condition at the peak of the pulsed field, this feedback control system is very beneficial for precise NMR measurements. In addition to the quasi-steady condition, which is preferred for NMR measurements, the high field reproducibility is equally important.
The field strength at the sample is continuously monitored and adjusted to the target value using an inner compensation coil, ensuring a high reproducibility of both the field strength and its time dependence. The quasi-steady field simplifies the NMR measurements, although the time at constant peak field is limited. In the flat-top regime, PFNMR measurements closely resemble those conducted in a steady field, allowing for the averaging of NMR spectra from different pulses to enhance the signal-to-noise ratio, and to combine several spectra to scan a broad frequency range. 
Employing the flat-top method enables conventional $T_1$ and $T_2$ measurements using standard pulse sequences as illustrated in Fig.~\ref{T2T1Meas}, significantly enhancing the experimental capabilities in PFNMR studies.

\subsection{NMR setup for pulsed-field facilities}

A time-dependent external magnetic field leads to a time-dependent Larmor frequency of the nuclear magnetic moments. Under these conditions, an NMR signal can be observed if the time-dependent field does not drive the spectral components outside the experimental bandwidth during the timescale of the NMR measurement, which is 10 to 100 microseconds, typically.
As described in Section 1, in an FID NMR experiment, a $\pi$/2 pulse of a few microseconds excites the nuclear spin ensemble. The FID signal appears after this RF pulse, decaying on a timescale of typically tens of microseconds. If the external magnetic field, and thus the Larmor frequency, changes by an amount comparable to the intrinsic NMR linewidth during this time, the oscillation of the observed FID signal will be frequency modulated, necessitating a demodulation analysis as explained in the next subsection.

\begin{figure}[tbp]
	\centering
	\includegraphics[width=0.8\linewidth]{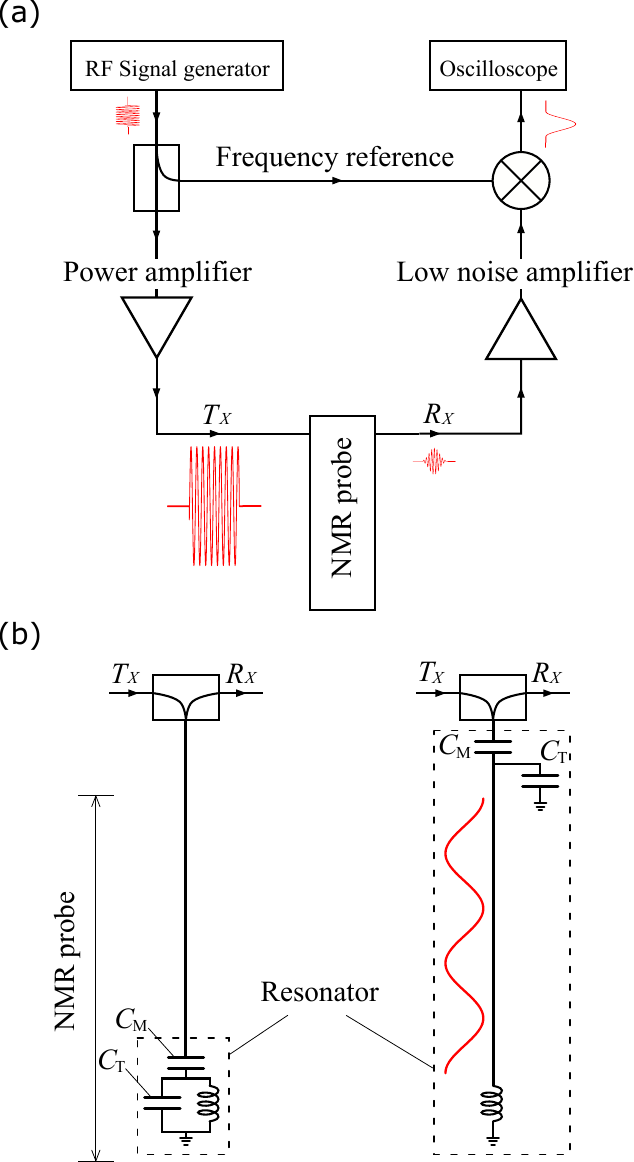}
	\caption{\label{nmrprobe} (a) Block diagram of an NMR spectrometer. The excitation pulse is amplified and transmitted to the NMR probe, and the detected response is then amplified and downconverted to the baseband for oscilloscope recording. (b) NMR probe design for bottom-tune (b, left) and top-tune (b, right) configurations. Two variable capacitors, $C_{\rm T}$ and $C_{\rm M}$, are used to tune the resonant frequency to the NMR measurement frequency and match the impedance to 50 ohms. In the bottom-tune configuration, the RF resonator is compactly positioned near the sample to optimize signal sensitivity and bandwidth. The top-tune configuration proves highly practical, as it enables resonance across a broad frequency spectrum using higher-order harmonics. In this setup, the RF signal propagates through the NMR probe, forming nodes, as represented by the red wave. With only a coil required inside the magnet at extreme conditions of high magnetic fields and often low temperatures, the top-tune configuration ensures highly stable experimental conditions.}
\end{figure}

A simple block diagram of an NMR experiment is presented in Fig.~\ref{nmrprobe}(a). The excitation pulse generated by the RF signal generator is amplified to a pulse power of up to several hundred watts by a power amplifier and transmitted to the NMR probe. The NMR signal, detected by the probe, is transmitted to a low-noise amplifier. The amplified NMR signal is then mixed with the reference frequency, split from the output of the signal generator, downconverted to the baseband and finally recorded by an oscilloscope. 

A schematic of NMR probes is shown in Fig.~\ref{nmrprobe}(b). To accurately record the NMR signal under transient conditions, the overall frequency bandwidth of the NMR detection must be sufficiently wide. The excitation frequency bandwidth for the transmitted pulse generating the RF field, often referred to by the technical term $T_X$ (compare Fig.~\ref{nmrprobe}), is mainly determined by the duration of the RF pulse.  
The frequency spectrum of the transmitted pulse results from the convolution of the square-shaped pulse waveform with the carrier frequency. 
The bandwidth is inversely proportional to the RF pulse duration. For example, to achieve a bandwidth exceeding 1 MHz, the RF pulse duration must be less than 1 $\mu$s. Shorter RF pulses require higher peak voltages to compensate for the RF power spread across a broader frequency range.  
Consequently, the transmitted pulse must be significantly amplified using a high-power amplifier, and the RF tank circuit should be designed to prevent voltage arcing.
In practice, a resonance can still be observed even if the $\pi/2$ pulse condition is not perfectly met, although this reduces the detected signal intensity.
Intentionally employing low RF power through the so-called small-tipping-angle method is advantageous for PFNMR because it allows for a rapid repetition of the signal observation and places lower power demands on the transmitted pulse. Further details on the small-tipping-angle method will be discussed below.

After transmission of the high-power RF pulse to excite the nuclear moments at the resonance frequency, the resulting NMR signal is detected by a receiver. Even though the frequency bandwidth of the receiver is usually broad, the overall experimental bandwidth may be constrained by the quality factor of the RF resonator. To generate the RF magnetic field during an NMR measurement, the sample is placed within a small coil, which forms a part of the tank circuit. The RF power is efficiently transmitted to the tank circuit when the characteristic impedance matches 50 ohms at the carrier frequency. 
For the signal detection, the RF tank circuit acts as an antenna to probe the weak NMR signal. A high quality factor of this antenna is crucial for sensitive detection, but also corresponds to a small frequency bandwidth. For optimal performance, the entire tank circuit should be constructed close to the sample using compact components. However, in PFNMR, the available space for the sample is typically less than 15 mm in diameter, and metallic components must be avoided in pulsed fields to prevent vibrations and heating from eddy currents. Consequently, a top-tuning configuration is often employed. In this configuration, only the sample within the RF coil is placed into the center of the pulsed-field coil.
As shown in Fig.~\ref{nmrprobe}(b, right), the RF coil is connected to a coaxial cable, and the variable capacitors for matching and tuning are located outside the pulsed magnet. This setup ensures a stable resonance under varying temperature and field conditions in the sample space. However, the frequency bandwidth in this configuration is narrower compared to the bottom-tuning configuration, with a typical frequency window of about 1 MHz.

During the pulsed field, particularly in the peak regime, the NMR pulse sequence is rapidly repeated to collect maximum data while the field strength matches the resonant conditions, typically within a timeframe of a few milliseconds.
Within a spin-echo sequence, two RF pulses are applied with a separation time shorter than $T_2$, the coherence time of the nuclear spin precession, compare Fig.~\ref{T2T1Meas}(b). On the other hand, if the repetition time between the pulse sequences is shorter than $T_2$, the evolution of the coherent nuclear spin state during one sequence overlaps into the next pulse sequence, resulting in undesirable signal correlation. Thus, a subsequent spin-echo sequence should be started only after a waiting period longer than $T_2$. Lastly, an accurate background subtraction requires that the interval between pulse sequences is long enough to observe the noise floor. 

In most solid-state materials, $T_2$ is much shorter than $T_1$. Therefore, when NMR measurements are performed at a fast repetition rate, the nuclear spins, once excited by the RF field and driven to decoherence, do not relax to thermal equilibrium within a magnetic-field pulse. 

A practical approach for observing many NMR signals within a single field pulse is the small-tipping-angle method. This technique allows for the detection of the NMR signal even when the RF pulse amplitude is insufficient to achieve a $\pi$/2 condition, allowing a part of the nuclear magnetization to contribute to the resonance signal in one RF pulse sequence, while preserving the remainder for the subsequent sequences.
This method is particularly effective for sweeping the magnetic field across a wide range to observe NMR signals at previously unknown resonance frequencies, or to record a broad NMR spectrum in several frequency steps. 
However, the amplitude of the detected signal is lower than that from a $\pi/2$ pulse, necessitating a sample with a strong NMR signal.

Another protocol for frequency-sweep spectrum measurements uses the flat-top region of a pulsed magnetic field. With a precisely controlled flat-top field, NMR measurements can be conducted in a manner similar to those in steady fields.
If the peak field is reproducible between subsequent field pulses within the precision limit provided by an auxiliary pick-up coil, the signal-to-noise ratio of the combined spectrum can be improved by repeating the NMR signal acquisition many times during multiple pulsed fields. However, the operational effort for repeating pulsed fields in this manner is rather high. 

To maximize efficiency, as many NMR spectra as possible should be recorded during the few milliseconds of the flat-top region within a single field pulse. 
However, since the flat-top duration is often shorter than the $T_1$ relaxation time, a $\pi / 2$ pulse can be applied only once for a given frequency regime of the spectrum, as defined by the experimental bandwidth. 
To conduct NMR measurements more frequently than $T_1$ permits, the center frequency must be shifted in steps of the experimental bandwidth between the RF pulses. Realizing rapid frequency shifts is technically challenging using analog circuits.
Therefore, digital-signal processing technology, such as software-defined radio (SDR), can be employed to facilitate rapid frequency shifts and to minimize the time required for hardware design and implementation.
SDR allows for the flexible generation of RF pulses by selecting appropriate waveform data, enabling the use of shaped pulses to enhance the power distribution efficiency in the frequency domain, as opposed to the usually employed power spectrum of a rectangular RF pulse. This well-defined excitation of specific parts of the NMR spectrum minimizes the impact on the remaining spectrum, which is subsequently excited by RF pulses at other center frequencies. The resulting NMR signals are then Fourier transformed and combined to obtain the complete frequency-domain spectrum. 

\begin{figure}[tbp]
	\centering
	\includegraphics[width=0.9\linewidth]{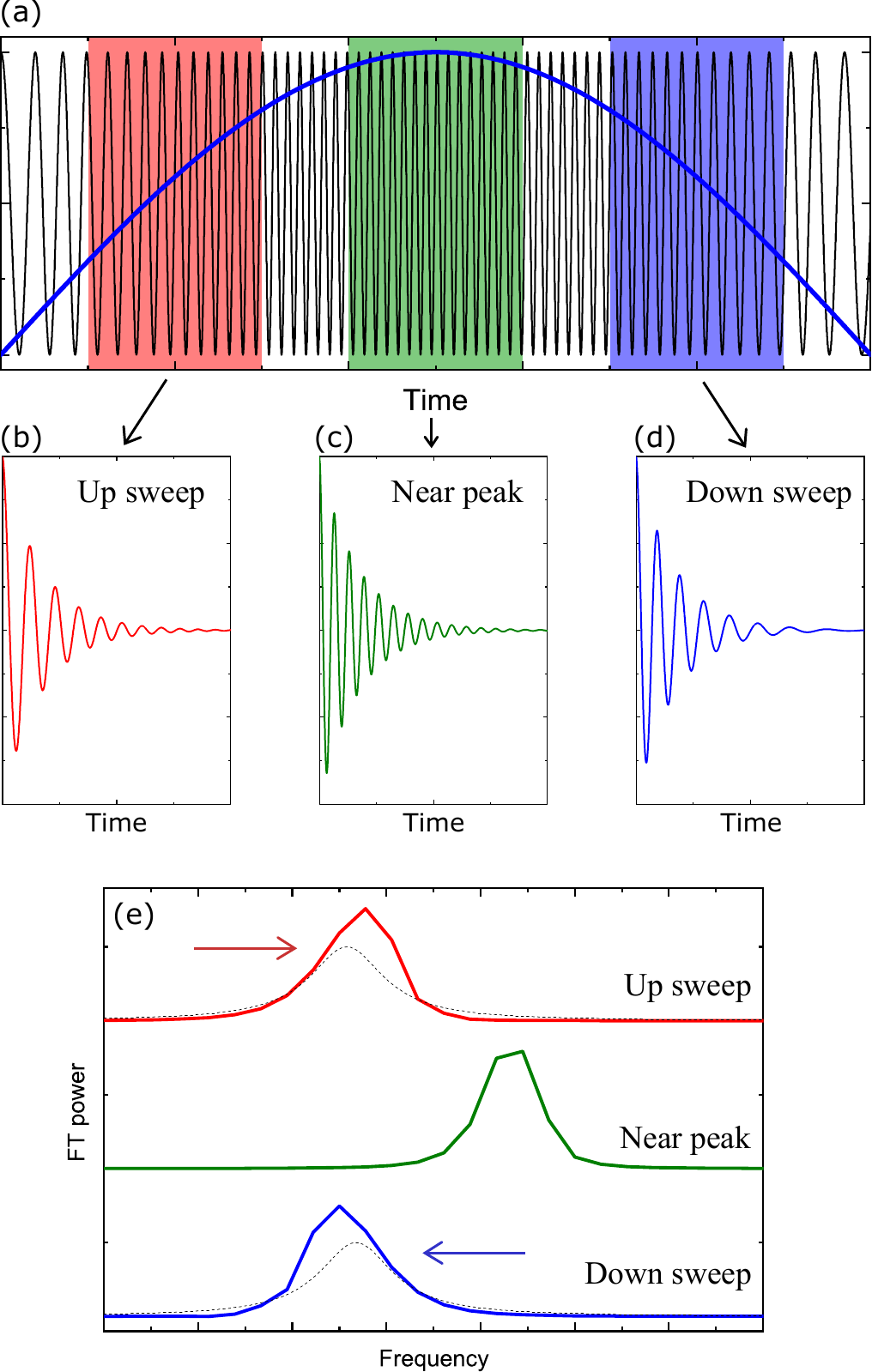}
	\\[2ex]
	\includegraphics[width=0.95\linewidth]{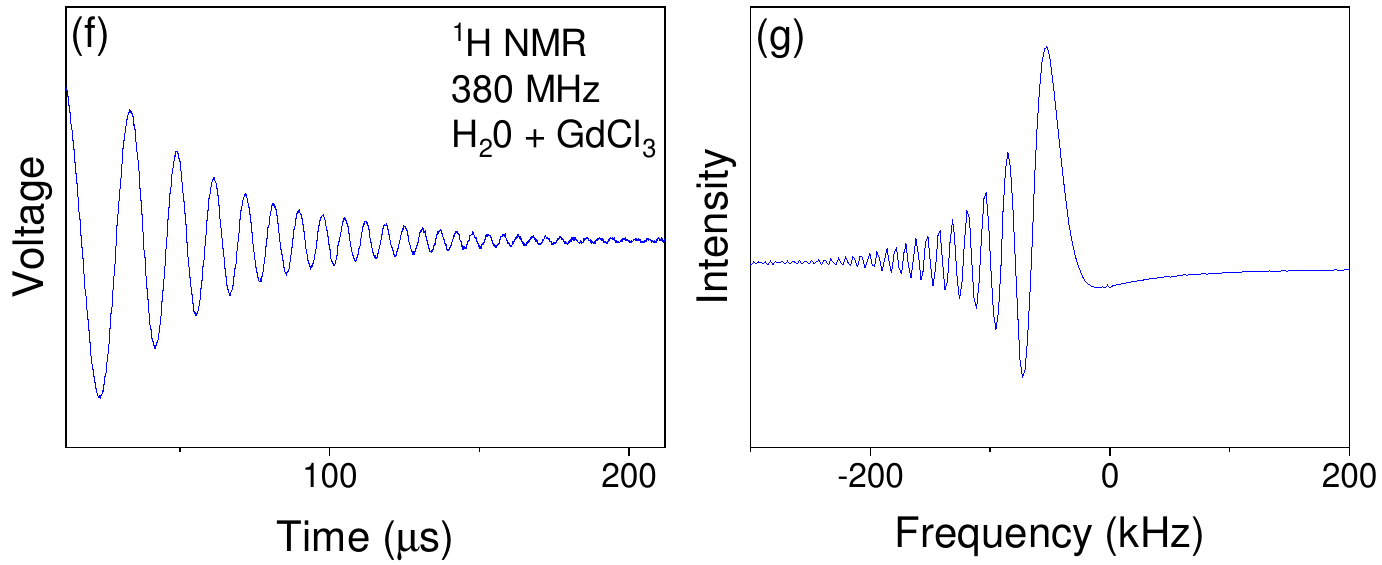}
	\caption{\label{fm} Effect of frequency modulation due to a time-dependent magnetic field. (a) Frequency modulation of a continuously oscillating simulated signal (black line), stemming from the time dependence of the magnetic field (blue). (b)-(d) FID signals obtained at characteristic times with corresponding colors. The displayed signals are multiplied by an exponential function to account for the $T_2$ decay. (e) Fourier transform of the FID signals. Dotted black lines indicate a symmetric Lorentzian function for comparison. When the external magnetic field increases (decreases), additional intensity is observed on the higher (lower) frequency side due to the frequency modulation. (f) Time-domain and (g) frequency-domain $^1$H NMR spectra of H$_2$0 with dissolved GdCl$_3$, recorded during the decrease of a pulsed field at a carrier frequency of 380 MHz. The spectrum in (g) exhibits an intrinsic linewidth of approximately 10 kHz, which is modulated and thus exhibits substantial inhomogeneous broadening due to the time-varying field.}
\end{figure}

\subsection{NMR data in pulsed fields and analyses procedure}

For PFNMR, the time-dependent variation of the pulsed magnetic field must be considered when analyzing the NMR spectrum. 
As the Larmor frequency $\omega (t)=\gamma H(t)$ changes with time in accordance with the pulsed magnetic field, the NMR signal is frequency modulated, similar to that in an FM radio. 
The modulated time-domain signal can be written as 
\begin{equation}
    f(t) = A\sin \left( \omega_0 t + \int_0^{t} \omega (\tau) d\tau \right),
\end{equation}
where the time evolution of the Larmor frequency $\omega(\tau)$ is included as an integral form in addition to the time-independent frequency $\omega_0$. 

Figure~\ref{fm}(a) shows the time-domain signal $f(t)$, modulated by the pulsed magnetic field, yielding an acceleration of the oscillation in the peak region. The oscillating signal can be considered in three regions: when the magnetic field is increasing (red), nearly constant (green), and decreasing (violet). 
In each region, when the FID signal is recorded, the resulting Fourier transformed spectrum deviates from a symmetric Lorentzian shape according to the change of magnetic field.
The FID signals illustrated in Figs.~\ref{fm}(b)--\ref{fm}(d) are modeled by multiplying an exponential decay function with the oscillating signal. The frequency spectra for each FID signal are shown in Fig.~\ref{fm}(e). 
The FT spectra obtained during rapid changes of the magnetic field (red and blue) exhibit an asymmetric spectral shape, highlighting the impact of the field variation during the recorded FID signal.  
Conversely, at the peak of the field, where the magnetic field changes only minimally, a more intrinsic and symmetric spectral lineshape is observed.
As the field increases (red peak), the peak frequency shifts to higher values, following the increase in the magnetic field. Conversely, during the decrease of the field (blue peak), the peak is broadened in the opposite direction.

The impact of a time-dependent magnetic field becomes significant when its variation during the acquisition of the FID signal is comparable to or larger than the intrinsic linewidth. The frequency modulation can be removed from the data by measuring the field profile $H(t)$ in various ways, for example, from the peak frequency of the FT spectra, using a second NMR signal as an internal reference, or via a pickup voltage. Subsequently, the data can be demodulated using the obtained $H(t)$. This is crucial for obtaining a sharp NMR spectrum. 

\begin{figure}[tbp]
	\centering
	\includegraphics[width=0.85\linewidth]{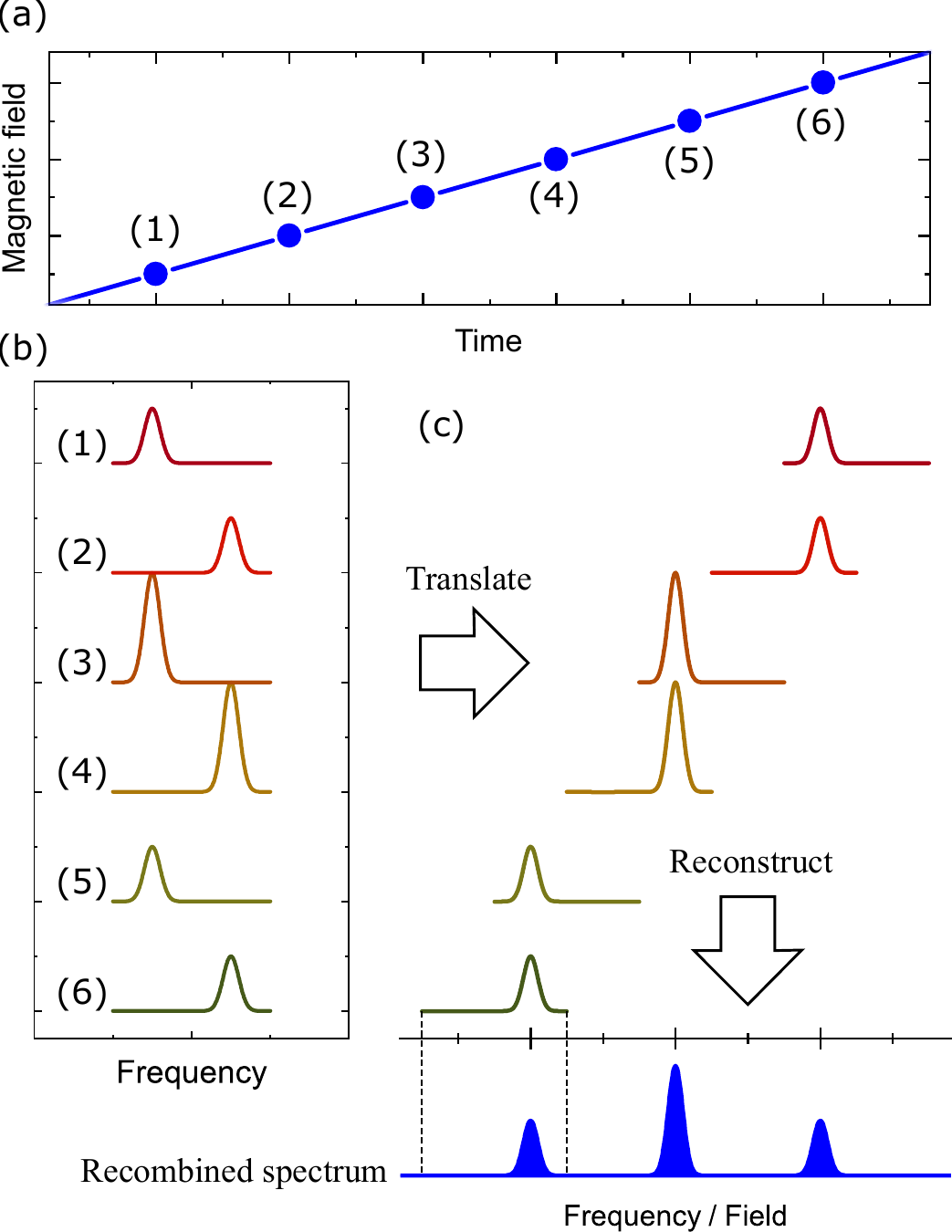}
	\\[2ex]
	\includegraphics[width=0.95\linewidth]{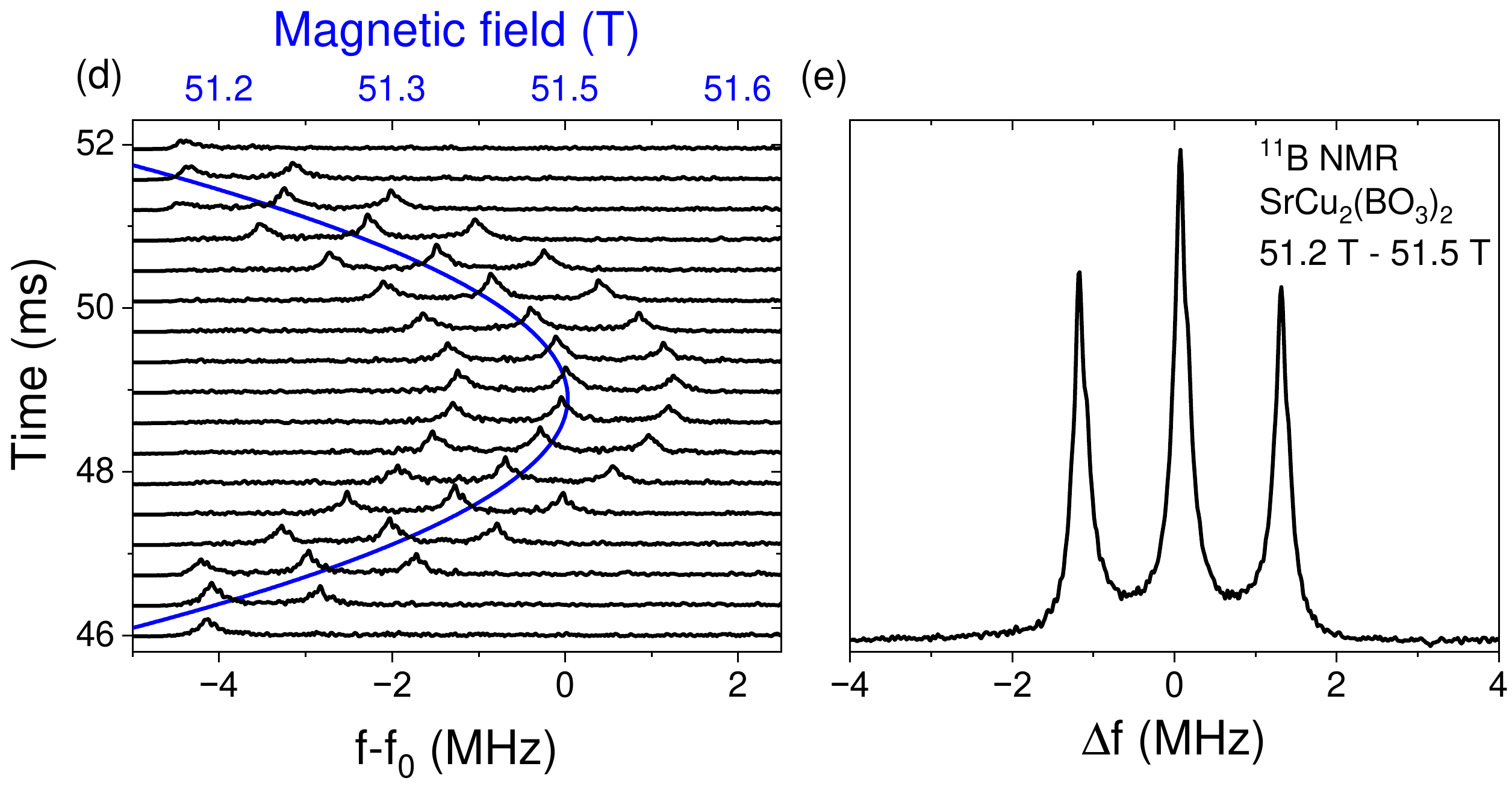}
	\caption{\label{recomb} Stepwise recording of broadband spectra. (a) Time-dependent magnetic field, assumed to linearly increase for simplicity. NMR spectra are recorded at each field point marked by numbers. (b) FT Spectra measured at the corresponding numbered points. Since the center frequency is constant, the peak frequency varies with the changing field strength. (c) The spectra are shifted corresponding to the external field at the time of measurement. By integrating all spectra on the field axis, the complete spectrum is obtained. (d) $^{11}$B NMR magnitude spectra of the material SrCu$_2$(BO$_3$)$_2$, recorded at the maximum region of a pulsed field, reaching up to about 51.5 T. (e) $^{11}$B NMR spectrum, resulting from shifting all spectra from (d) according to the time-dependent field and subsequent summation from about 51.2 to 51.5 T (unpublished). }
\end{figure}

In PFNMR experiments, broadband NMR spectra with spectral features that extend across a wide frequency range of several MHz can, similar to field-sweep procedures in conventional condensed-matter NMR experiments, be recorded in moderately fast changing magnetic fields by mapping the transient spectra onto the magnetic field axis.
Figure~\ref{recomb} illustrates this process. Here, six transient FT spectra are collected while the magnetic field increases at a constant rate. The different times of the FT spectrum acquisition are indicated by blue circles and indexed to trace the correspondence between the field and spectra for further data analysis. 
Each FT spectrum is confined within a frequency window set by the experimental bandwidth. 
Figure~\ref{recomb}(b) shows the resulting FT spectra, with frequencies relative to the fixed center frequency of the NMR measurement. 
Figure~\ref{recomb}(c) displays how these transient FT spectra are shifted according to the magnetic field at the time of measurement.
The frequency axis of the FT spectrum is converted to the field axis using the formula $\delta H/H_0 =-\delta \omega / (\omega_0+\delta \omega) $ around $\omega_0=\gamma H_0$.
The mapped FT spectra are then integrated along the frequency or field axis to produce the complete, broad NMR spectrum, represented by three blue spectral lines at the bottom of Fig.~\ref{recomb}(c).

As an example of experimental data, Fig.~\ref{recomb}(d) shows $^{11}$B NMR magnitude spectra of the material SrCu$_2$(BO$_3$)$_2$, recorded at the peak of a pulsed magnetic field reaching approximately 51.5 T at a sample temperature of 100 K. Fig.~\ref{recomb}(e) displays the resulting $^{11}$B NMR spectrum, obtained by shifting all spectra from (d) in accordance with the time-dependent field and subsequent summation. Due to this summation procedure, the resulting spectrum represents a range of fields from about 51.2 T to 51.5 T, rather than a single field strength. In such experiments, the number of FID spectra can reach several hundred during a single field pulse. To streamline the analysis, it is highly beneficial to develop custom software that aligns numerous spectra to the NMR center frequency and facilitates their conversion into the frequency or field domain based on the time-dependent pulsed-field data.

Measuring the nuclear relaxation times $T_1$ and $T_2$ in pulsed magnetic fields presents significant challenges due to the difficulty of achieving a thermal equilibrium state of the nuclear spin system. Here, using flat-top pulsed fields allows for conventional saturation-recovery methods to measure $T_1$, although some constraints are imposed by the limited duration of just a few milliseconds of the flat-top regime. On the other hand, $T_2$ measurements are more feasible using flat-top pulses, since the typical $T_2$ timescale is less than 1 ms.

However, it is also possible, though more complex, to measure relaxation times without a flat-top pulse. For example, $T_1$ measurements of aluminum powder and liquid gallium have been performed in regular pulsed fields \cite{Kohlrautz2016}.
Here, to achieve $T_1$ results consistent with those obtained in steady fields, an inversion RF pulse is applied just before the peak field, followed by recording the NMR signal intensity using low-power RF pulses after the inversion.
As the excitation pulse tilts the nuclear moments by only a small angle, the progressive recovery of the NMR signal intensity can be monitored during the variation of the pulsed field around the resonant condition.
The obtained recovery curve can then be fitted to a theoretical model that considers several experimental parameters, such as the tipping angle of the RF pulses and the properties of the tank circuit.

For $T_2$ measurements, a Carr-Purcell pulse sequence can be utilized \cite{Meier2012}. 
Here, following the initial $\pi$ pulse, further $\pi$ pulses are applied, causing the dephasing nuclear moments to refocus at $2t_{\rm pp}$, and subsequently at every $2t_{\rm pp}$ interval. The signal intensity decays on a timescale of $T_2$ after several refocusing events.
Notably, the measured $T_2$ is significantly longer than the decay constant of an FID, $T_2^\ast$, which arises from the spatial distribution of magnetic fields. Furthermore, external magnetic field fluctuations during the Carr-Purcell sequence can lead to a shortening of $T_2$. For optimal results, the Carr-Purcell pulse sequence should be employed in the flat-top regime of a pulsed field. 

\begin{figure}[tbp]
	\centering
	\includegraphics[width=0.90\linewidth]{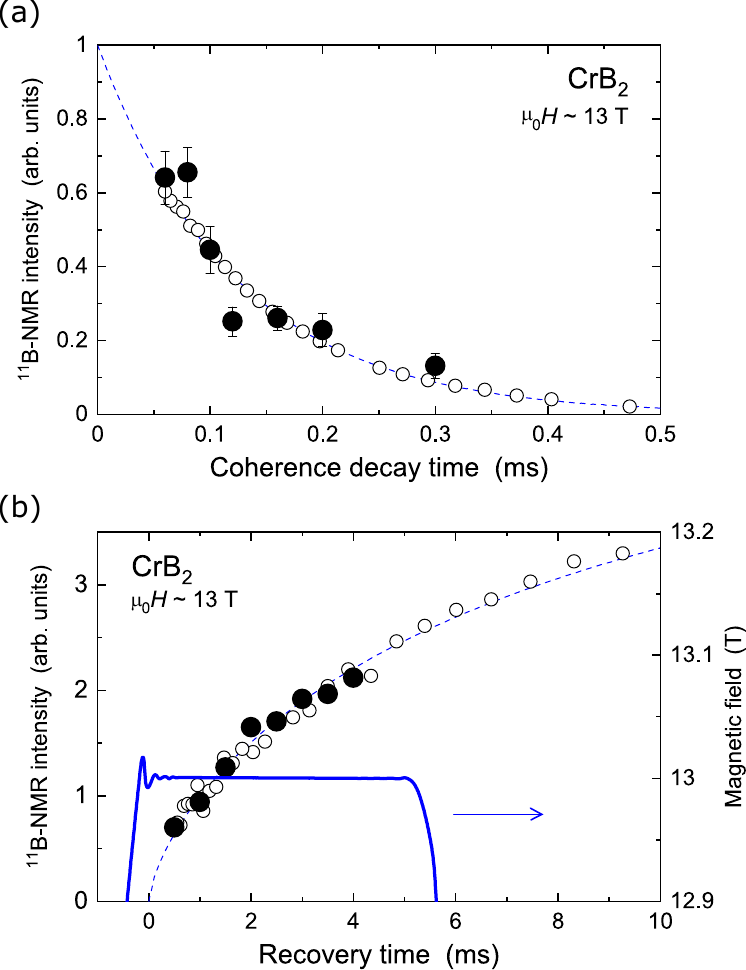}
	\caption{\label{T1T2_CrB2} Results of (a) $T_2$ and (b) $T_1$ measurements of CrB$_2$ in a flat-top pulsed field (black points). Open symbols indicate the results obtained in steady fields for comparison. The blue solid line in (b) represents the field profile of the flat-top pulsed field. The field is kept constant during the recovery-curve measurement. Due to the limited duration of the flat-top pulse, only the initial part of the $T_1$ recovery curve is obtained.}
\end{figure}

The relaxation-time measurements are significantly simplified when performed in the quasi-steady regime of a flat-top pulsed field. 
Figures~\ref{T1T2_CrB2}(a) and \ref{T1T2_CrB2}(b) present the $T_2$ decay and $T_1$ recovery curves for CrB$_2$, measured in a flat-top regime \cite{Ihara2021}. Since the magnetic field is constant during the time of the $T_2$ measurement, a conventional $T_2$ pulse sequence, as illustrated in Fig.~\ref{T2T1Meas}(a)--\ref{T2T1Meas}(c), can be employed. 
In order to measure the full $T_1$ relaxation, the duration of the flat part should be several times longer than $T_1$.
The pulsed magnet used for the $T_1$ measurements of CrB$_2$, shown in Fig.~\ref{T1T2_CrB2}(b), provides a flat-top duration of about 4 ms, allowing only the initial part of the recovery curve to be measured (black points). Therefore, a longer pulse duration would be necessary to measure the entire recovery curve.

\section{3. Exemplary applications}

The research of materials with complex interactions and strong correlations of the electronic degrees of freedom is a very broad field of research in contemporary physics, encompassing both fundamental research topics as well as  applications of new materials. 

The family of these materials includes, for example, metallic, insulating, as well as organo-metallic or purely organic compounds, either in single-crystalline or polycrystalline form.
Particularly in materials of substantial significance for both fundamental research and industrial applications, the energy scale of the electronic interactions is often of the order of several meV, corresponding to Zeeman energies for fields of several tens of teslas. A prominent example, first reported in 1986 by Bednorz and M\"{u}ller, are the cuprate high-temperature superconductors, which feature layers of copper oxides alternating with layers of other metal oxides \cite{Bednorz1986}. More recently, the critical temperatures for superconductivity in a very different class of materials, namely highly-pressurized hydrides, have been reported to exceed 200 K \cite{Livas2020}. 

Materials with complex and frustrated magnetic interactions may exhibit unique ground states with field-driven magnetic order and exotic many-body excitations in high magnetic fields. This includes the study of Mott insulators, which embody fundamental models of low-dimensional spin lattices, such as spin chains, spin ladders, and extended quasi-two-dimensional systems \cite{Schollwoeck2004}.
In these systems, fundamental predictions of theoretical models can be tested, but also novel and exotic ground states and excitations of collective electron systems may be found. In this context, PFNMR has numerous applications and can provide key insights for a fundamental understanding of topical phenomena.

In the following, we will present and discuss several selected examples of PFNMR studies from the literature.

\subsection{Hydrogen NMR at GHz frequencies}

\begin{figure}[tbp]
	\centering
	\includegraphics[width=0.8\linewidth]{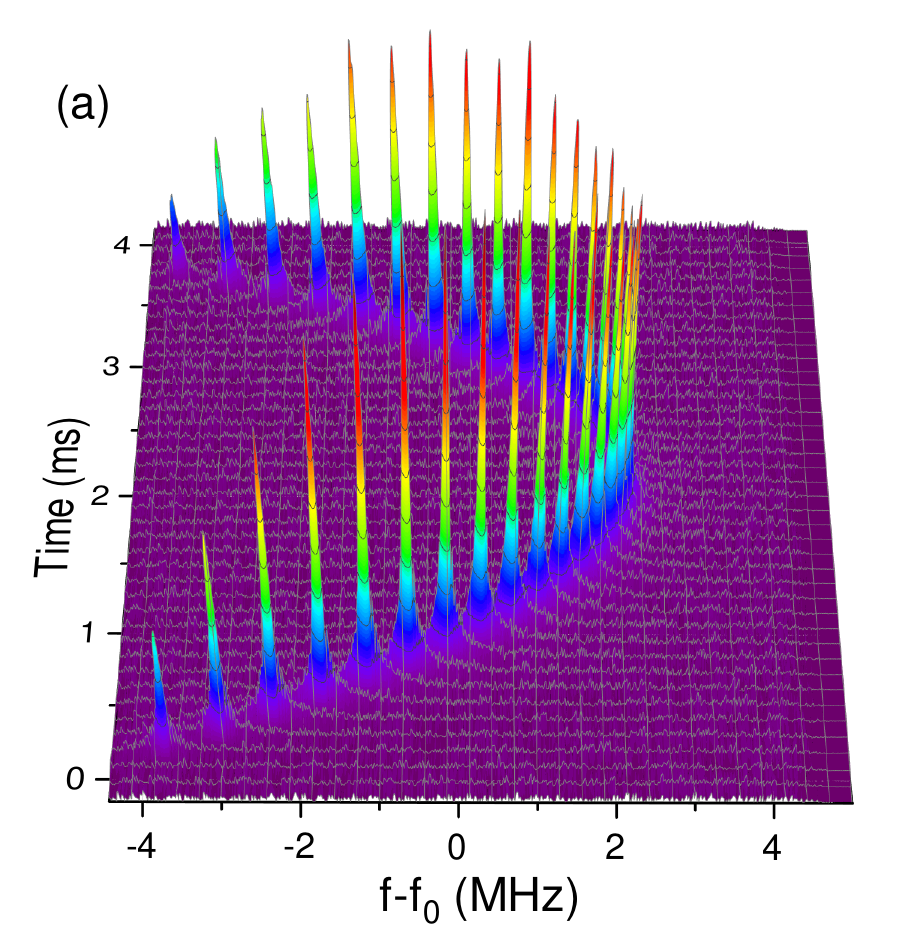}
	\\[2ex]
	\includegraphics[width=0.7\linewidth]{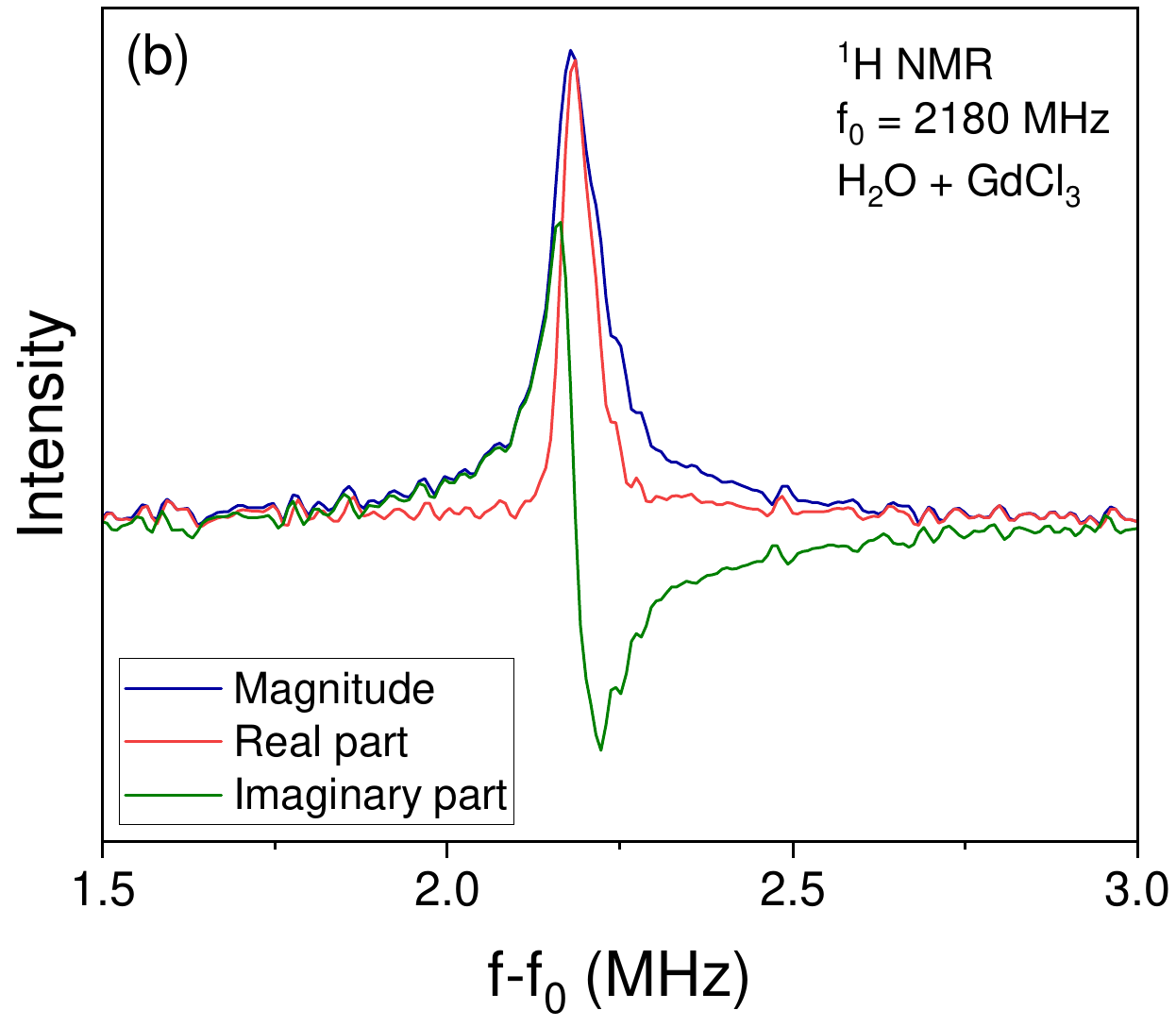}
	\caption{\label{1H_2GHz} (a) 3D contour plot of frequency-domain $^1$H NMR spectra of H$_2$0 with GdCl$_3$ dissolved in it, recorded at a center frequency of $f_0 = 2180$ MHz and room temperature. (b) $^1$H NMR spectrum at the pulsed-field maximum of approximately 51.2 T in (a) with a resonance frequency of 2182.178 MHz (unpublished).}
\end{figure}

The highest observable hydrogen resonance frequency is a crucial benchmark for any NMR apparatus. Hydrogen, as the most commonly studied nuclear isotope in NMR spectroscopy, is predominantly utilized due to its abundance in organic molecules and its strong signal intensity in NMR experiments.

An accurate probing of the $^1$H NMR frequency is essential for obtaining highly-resolved spectral information. It facilitates the precise determination of chemical shifts, coupling constants, and relaxation times, which are integral for elucidating molecular structures and dynamics. Moreover, the $^1$H NMR frequency serves as a standard reference, ensuring that NMR results are reproducible and comparable across different experimental setups.

One application of PFNMR in this context is the synthesis and testing of new contrast agents that are specifically designed for use in high-field MRI scanners \cite{Din2023}. These contrast agents are required to maintain a high relaxivity at elevated frequencies, allowing for improved image quality and sensitivity at increased magnetic fields.
In addition to the creation of new contrast agents, optimizing their formulation and delivery methods is crucial. This optimization ensures efficient distribution within the body and targeted accumulation at specific sites, potentially involving encapsulation within nanoparticles or other carriers to improve stability and biodistribution.

High-frequency NMR experiments present specific challenges, primarily due to increased RF signal attenuation in cables and other components of the NMR spectrometer. This attenuation can result in signal loss and degradation, compromising the quality of the acquired spectrum.
To address this issue, several strategies are employed. High-quality cables with low attenuation characteristics are utilized to minimize signal loss during transmission. Additionally, the careful design of RF coils and other components helps to optimize the signal-transmission efficiency. Furthermore, pre-amplifiers located close to the RF coils can amplify weak signals before they undergo attenuation below the noise floor.

Despite these instrumental challenges, it is possible to record high-quality $^1$H NMR spectra in pulsed magnetic fields with resonance frequencies up to the GHz regime. As an example, Fig.~\ref{1H_2GHz}(a) shows a series of $^1$H small-tip-angle FID spectra, recorded around the peak of a magnetic-field pulse up to approximately 51.2 T. RF pulses with a duration of 0.2 $\mu$s were applied at a center frequency of 2180 MHz and with a repetition time of 100 $\mu$s. Figure \ref{1H_2GHz}(b) shows a spectrum recorded at the maximum of the field pulse, yielding a full width at half maximum (FWHM) linewidth of about 50 kHz and a signal-to-noise ratio of 30.

\subsection{Spin nematicity in LiCuVO$_4$}

\begin{figure}[tbp]
	\centering
	\includegraphics[width=0.9\linewidth]{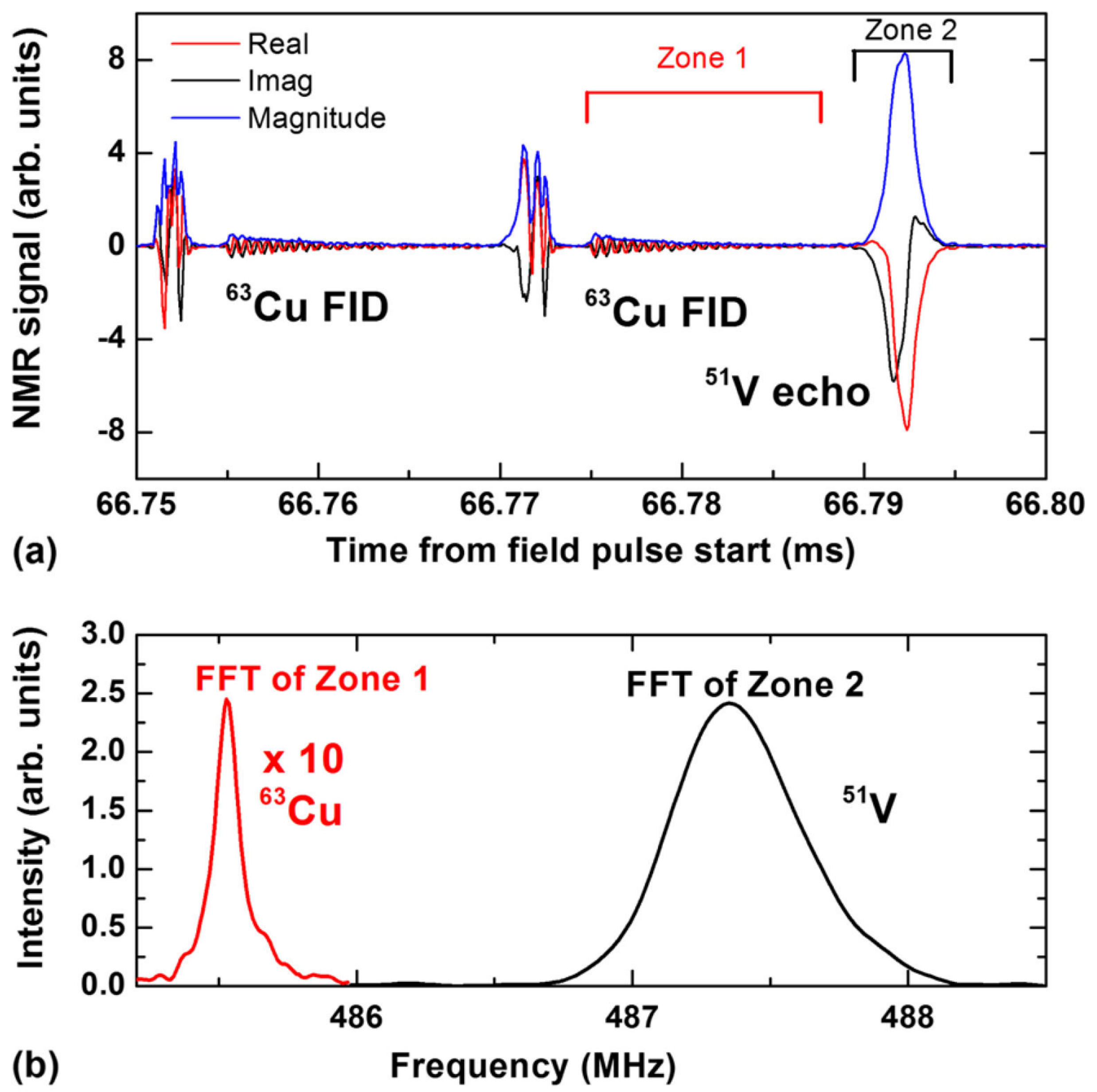}
	\caption{\label{LiCuVO_Fig1} (a) Simultaneous NMR time record of the $^{63}$Cu-metal FID and $^{51}$V spin echo of LiCuVO$_4$ for $H \parallel c$ at $\mu_0 H = 42.91$ T and a resonance frequency of 487.2 MHz. The two $^{63}$Cu FID signals are preceded by the strong transients from RF pulses, which saturate the receiver. (b) Fourier transforms of the NMR time-domain signals, applied separately to zone 1 and zone 2, to provide the NMR spectra of $^{63}$Cu and $^{51}$V, used for field reference and the determination of the local field in LiCuVO$_4$, respectively. Reproduced with permission from Orlova \textit{et al.}, Phys. Rev. Lett. \textbf{118}, 247201 (2017).  Copyright 2017 American Physical Society \cite{Orlova2017}.}
\end{figure}

In the study of low-dimensional quantum magnetism, the spin-nematic phase represents a particularly fascinating state, wherein the behavior of a quantum magnet resembles that of a liquid crystal. This phase is characterized by the absence of transverse dipolar order under an external magnetic field, while exhibiting transverse quadrupolar order instead.
The quadrupolar order parameter develops on the bonds between neighboring spins and can be conceptualized as a condensate of two-magnon pairs. It breaks the spin-rotational symmetry about the magnetic field, albeit only partially, as $\pi$ rotations transform the order parameter into
itself. Additionally, the translational symmetry of the order parameter is also broken, although this is not observable in the dipolar channel.

Employing $^{51}$V PFNMR, Orlova \textit{et al.}
characterized the microscopic high-field properties of the material LiCuVO$_4$ for
two crystal orientations, $H \parallel c$ and $H \parallel b$, see Ref.~\cite{Orlova2017} and citations therein. 
As shown in Figs.~\ref{LiCuVO_Fig1}(a) and \ref{LiCuVO_Fig1}(b), the NMR signal was recorded using echo-pulse sequences comprising two 0.5 $\mu$s excitation pulses separated by 20 $\mu$s, resulting in a spectral excitation bandwidth of 1.2 MHz. This is sufficient to record the entire $^{51}$V spectrum in a single data acquisition.
To determine the instantaneous value of the pulsed field, the $^{63}$Cu NMR signal from copper powder, placed in the NMR coil together with the single-crystalline sample of LiCuVO$_4$, was utilized. Because of the similiar gyromagnetic ratios of the $^{51}$V ($^{51}\gamma = 11.199$ MHz/T) and $^{63}$Cu ($^{63}\gamma = 11.285$ MHz/T) nuclei, each NMR time spectrum contains both the $^{51}$V spin-echo signal of LiCuVO$_4$ and the $^{63}$Cu FID signal of the copper powder. 

\begin{figure}[tbp]
	\centering
	\includegraphics[width=0.95\linewidth]{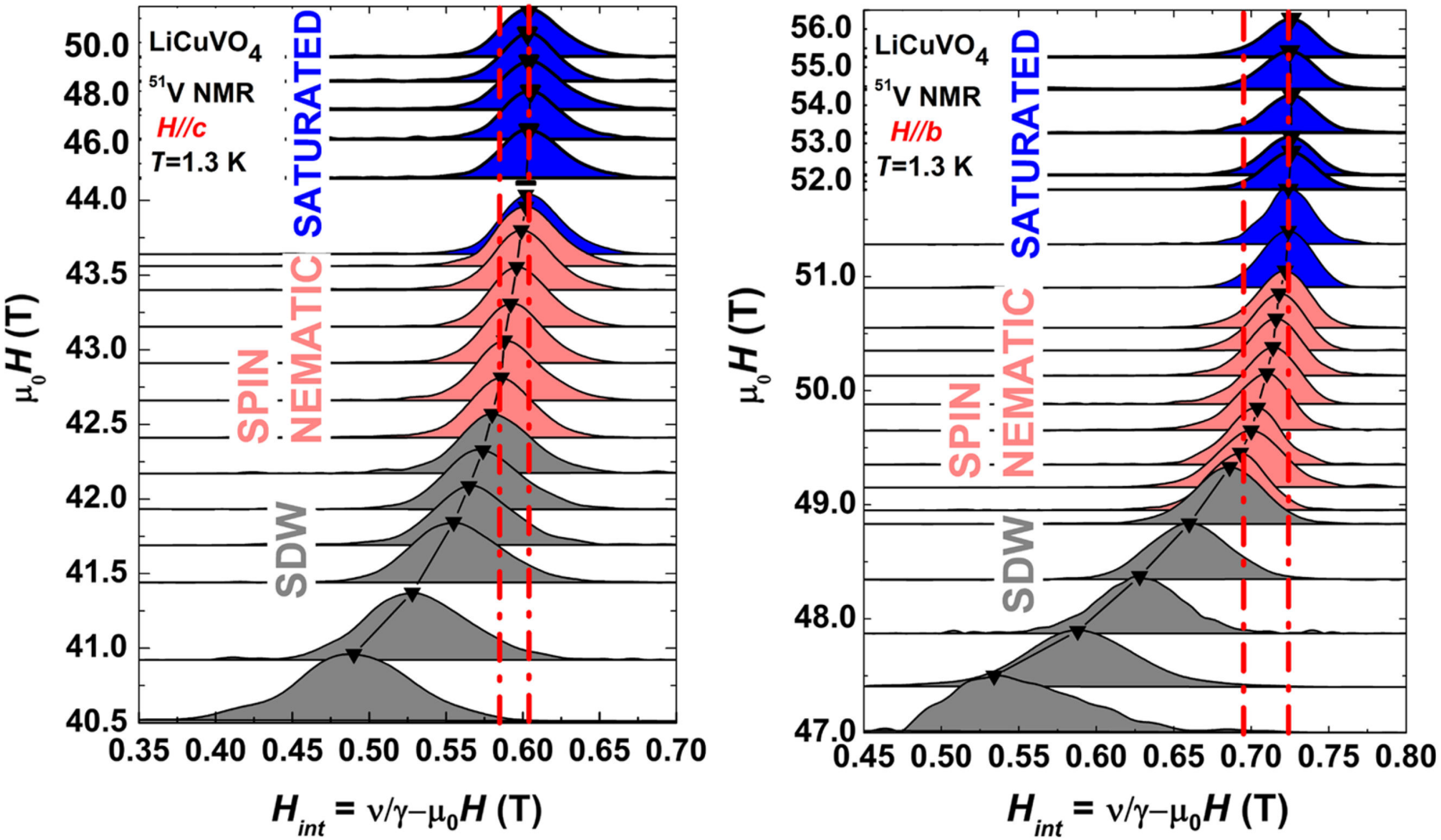}
	\caption{\label{LiCuVO_Fig2} Field dependence of the $^{51}$V NMR spectra of LiCuVO$_4$ for $H \parallel c$ (left) and $H \parallel b$ (right) at 1.3 K, normalized to their peak intensity. The black triangles mark the peak of each NMR line, demonstrating their shift towards the saturated state. Three different regions are identified: saturated (blue), spin-density wave (gray), and spin nematic (light red). The dash-dotted red lines denote the region where the internal field $H_{int}$ becomes field-dependent, but maintains the same distribution as in the saturated state. Reproduced with permission from Orlova \textit{et al.}, Phys. Rev. Lett. \textbf{118}, 247201 (2017).  Copyright 2017 American Physical Society \cite{Orlova2017}.}
\end{figure}

The NMR spectra of LiCuVO$_4$, recorded for a broad field range and orientations $H \parallel c$ and $H \parallel b$, are shown in Fig.~\ref{LiCuVO_Fig2}.
Just below full magnetic saturation, the $^{51}$V NMR spectra evidence a field range where a homogeneous local magnetization is increasing with field. The authors report that such behavior corresponds to the predicted spin-nematic phase. The experimentally observed field dependencies of spectral features, and their difference for $H \parallel c$ and $H \parallel b$, likely indicative of the easy-plane anisotropy in LiCuVO$_4$, provide a basis for further theoretical investigations of this material, and should help in distinguishing different models for the spin-nematic phase.

\subsection{The spin-dimer
antiferromagnet SrCu$_2$(BO$_3$)$_2$}

\begin{figure}[tbp]
	\centering
	\includegraphics[width=0.95\linewidth]{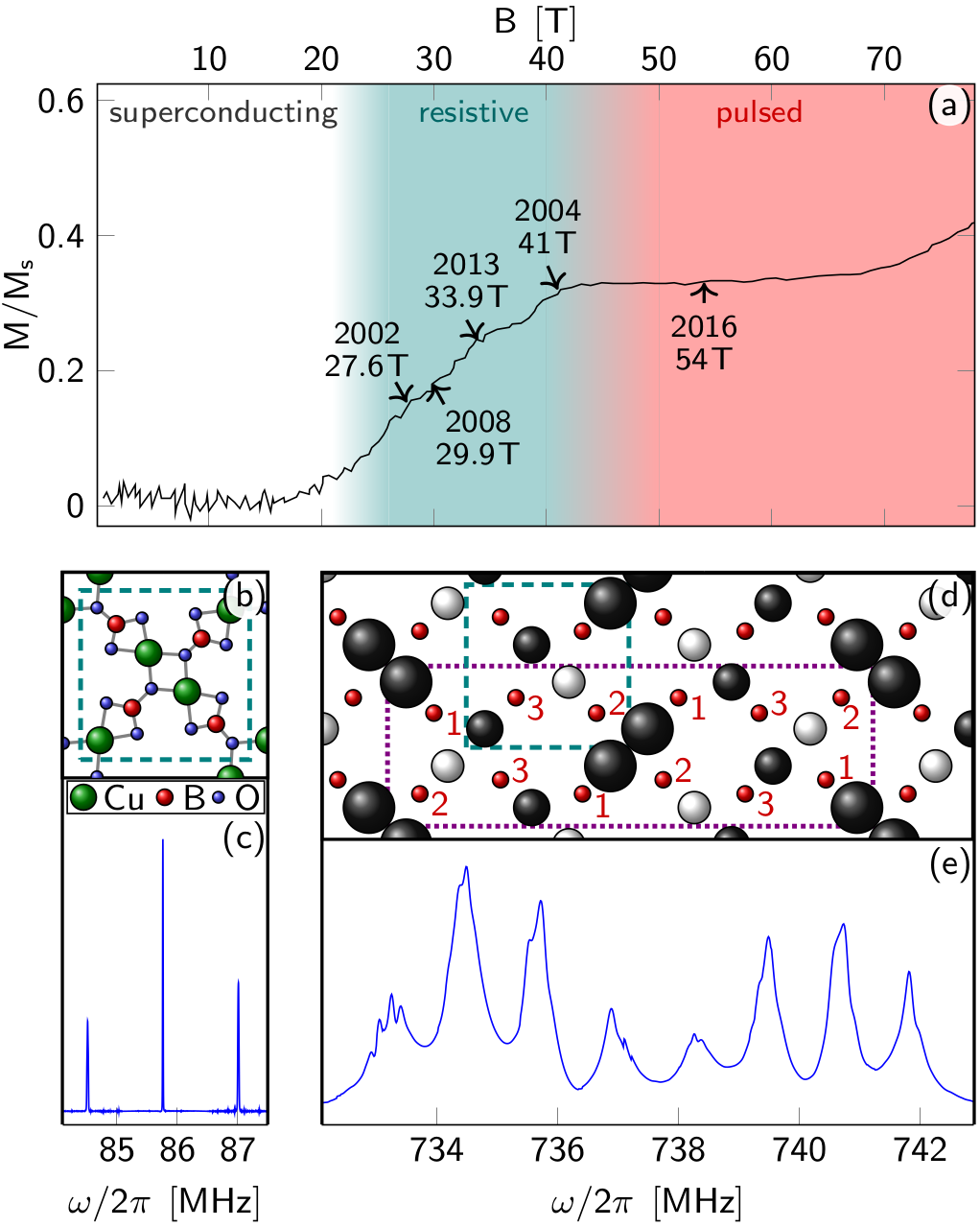}
	\caption{\label{Fig_SCBO} (a) Magnetization of SrCu$_2$(BO$_3$)$_2$ for $H \parallel c$ at 2.1 K. For previous NMR studies up to 41 T, superconducting and resistive magnets have been used (Ref. \cite{Kohlrautz2016SCBO} and citations therein). (b) Unit cell of the Cu$_2$(BO$_3$)$_2$ plane (green dashed box) and (c) the corresponding $^{11}$B NMR spectrum at 6.281 T and 5 K. (d) Magnetic superlattice at the 1/3 magnetization plateau (purple dotted box) and three inequivalent $^{11}$B sites in the plane (red spheres). White spheres and black spheres represent negative and positive electronic spin polarization, respectively. (e) $^{11}$B PFNMR spectrum at about 54 T and 2 K. Reproduced with permission from Kohlrautz \textit{et al.}, J. Magn. Reson. \textbf{271}, 52 (2016). Copyright 2016 Elsevier \cite{Kohlrautz2016SCBO}.}
\end{figure}

In this example, $^{11}$B PFNMR is used to study the spin-dimer
antiferromagnet SrCu$_2$(BO$_3$)$_2$ at 54 T and low temperatures, see Ref.~\cite{Kohlrautz2016SCBO} and citations therein. This material is an excellent realization of the Shastry-Sutherland lattice model, wherein the electronic spins of Cu$^{2+}$ ions within the Cu$_2$(BO$_3$)$_2$ layers arrange into a structure of mutually orthogonal spin-singlet dimers, exhibiting strong interdimer interactions that induce magnetic frustration. The ground state of the Shastry-Sutherland model is described by a product of spin-singlet dimer states, resulting in zero net magnetization at low temperatures. At elevated magnetic fields, triplet states with reduced kinetic energy condense, resulting in a sequence of field-driven magnetic superlattices, which manifest as plateaus of the macroscopic magnetization. The determination of the local spin polarization within these superlattice structures by means of NMR spectroscopy is of great interest in the research of SrCu$_2$(BO$_3$)$_2$.

The broken symmetries of the plateau-specific magnetic superlattices lead to complex NMR spectra, previously investigated using steady-field NMR up to 41 T, which is at the lower field end of the plateau with 1/3 of the saturation magnetization, see Fig.~\ref{Fig_SCBO}(a). At this plateau, the commensurate magnetic unit cell is enlarged by a factor of 3, see Fig.~\ref{Fig_SCBO}(d), and the hyperfine fields yield a spectrum that is spread over several MHz, see Fig.~\ref{Fig_SCBO}(e).

\begin{figure}[tbp]
	\centering
	\includegraphics[width=0.8\linewidth]{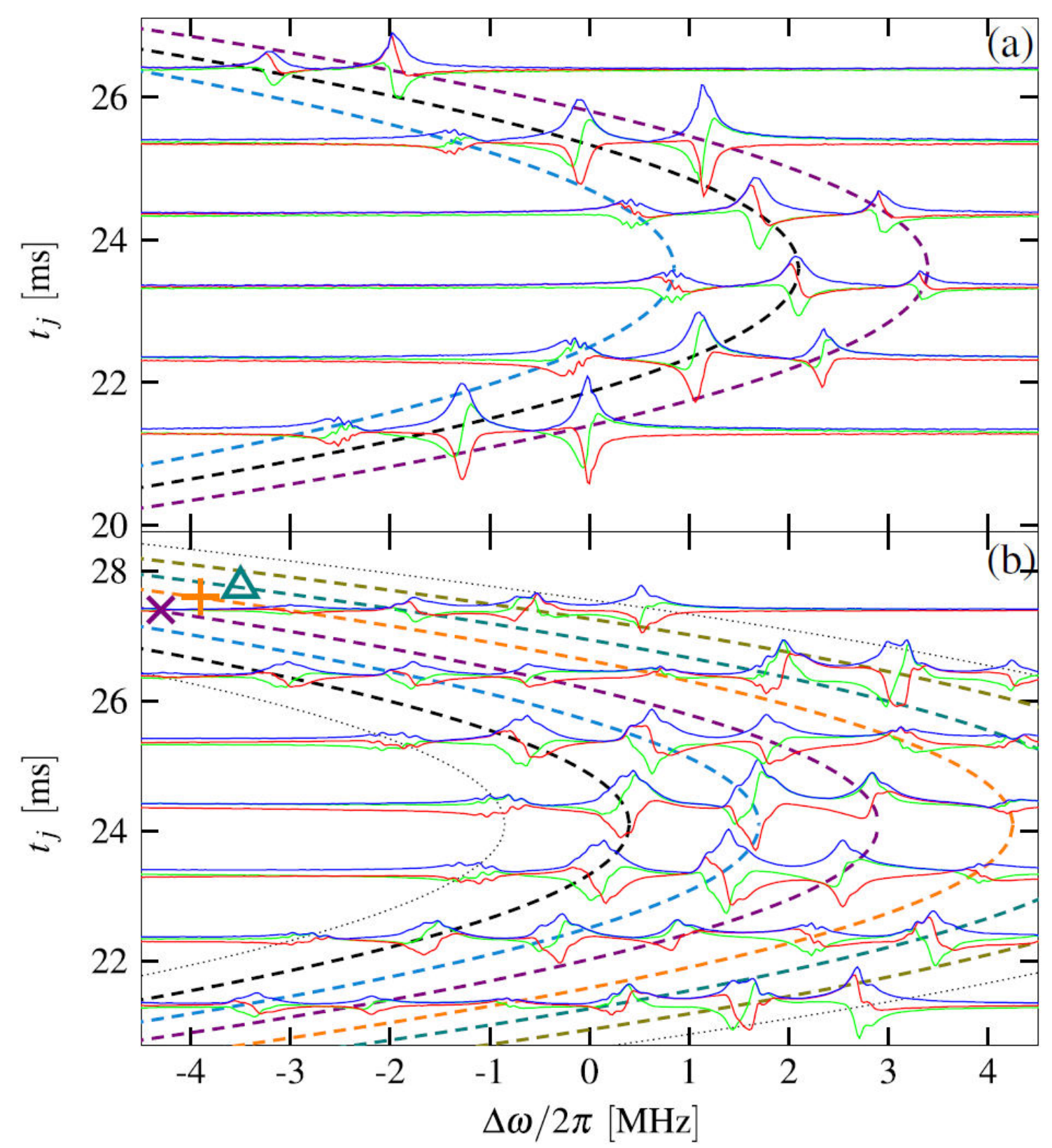}
	\caption{\label{Fig_SCBO_spectra} Magnitude (blue), real (red), and imaginary part (green) of the $^{11}$B PFNMR spectra of SrCu$_2$(BO$_3$)$_2$, recorded at different times $t_j$ at a field of about 54 T at (a) 119 K and (b) 2 K. Frequency offsets $\Delta \omega$ are defined relative to the carrier frequency of the RF pulses. Dotted/dashed curves show the approximate time-dependent field for each resonance line. Reproduced with permission from Kohlrautz \textit{et al.}, J. Magn. Reson. \textbf{271}, 52 (2016). Copyright 2016 Elsevier \cite{Kohlrautz2016SCBO}.}
\end{figure}

The $^{11}$B PFNMR FID spectra were recorded in the peak region of a pulsed magnetic field, oriented parallel to the crystallographic $c$ axis. Since the field maximum was not known with high enough precision prior to the experiment, a sequence of RF pulses with a fixed frequency of 740 MHz, a duration of 0.2 $\mu$s duration each, and with a delay of 500 $\mu$s was employed. A pulse power of about 50 W was chosen to ensure that an on-resonance pulse would reduce the longitudinal nuclear magnetization by less than 2 $\%$.

Representative results of the obtained $^{11}$B PFNMR spectra are shown in Fig.~\ref{Fig_SCBO_spectra}. Since the $^{11}$B isotope has a nuclear spin of $I = 3/2$, the spectrum at 119 K in the paramagnetic phase consists of three resonance peaks, stemming from the quadrupole interaction. The time dependence of the field is indicated by the dashed curves for each resonance line. The central and satellite transitions can be identified, and an intensity drop for high-frequency offsets due to the limited excitation bandwidth of the RF pulses is observed. 
Based on the central transition and the high-frequency satellite transition, a quadrupole splitting of 1.24 MHz is deduced, which is in good agreement with expectations from steady-field experiments. Furthermore, the two satellite peaks are not identical in shape. This can be attributed to a variation of the local electric field gradient and magnetic hyperfine shift tensors over the different $^{11}$B sites in the crystallographic unit cell of SrCu$_2$(BO$_3$)$_2$. Any small misorientation between the axis of the pulsed magnetic field and the crystallographic $c$ axis will break the symmetry of the satellites.

At 2 K, the emergence of the magnetic superstructure results in a complex distribution of static hyperfine fields, probed by NMR at the atomic $^{11}$B positions, as shown in Fig.~\ref{Fig_SCBO_spectra}(b). The stacking of the Cu$_2$(BO$_3$)$_2$ layers further increases the number of inequvalent $^{11}$B sites, and the overlap of the different quadrupolar-split, site-specific peaks requires a sophisticated deconvolution procedure to elucidate the magnetic structure. The observed spectrum closely resembles that previously reported at a steady field of 41 T and at 2 K, indicating a very similar distribution of hyperfine fields. These results establish the efficacy of PFNMR in studying complex magnetic order in bulk materials at the highest magnetic fields.

\subsection{The heavy-fermion compound CeIn$_3$}

\begin{figure}[tbp]
	\centering
	\includegraphics[width=0.95\linewidth]{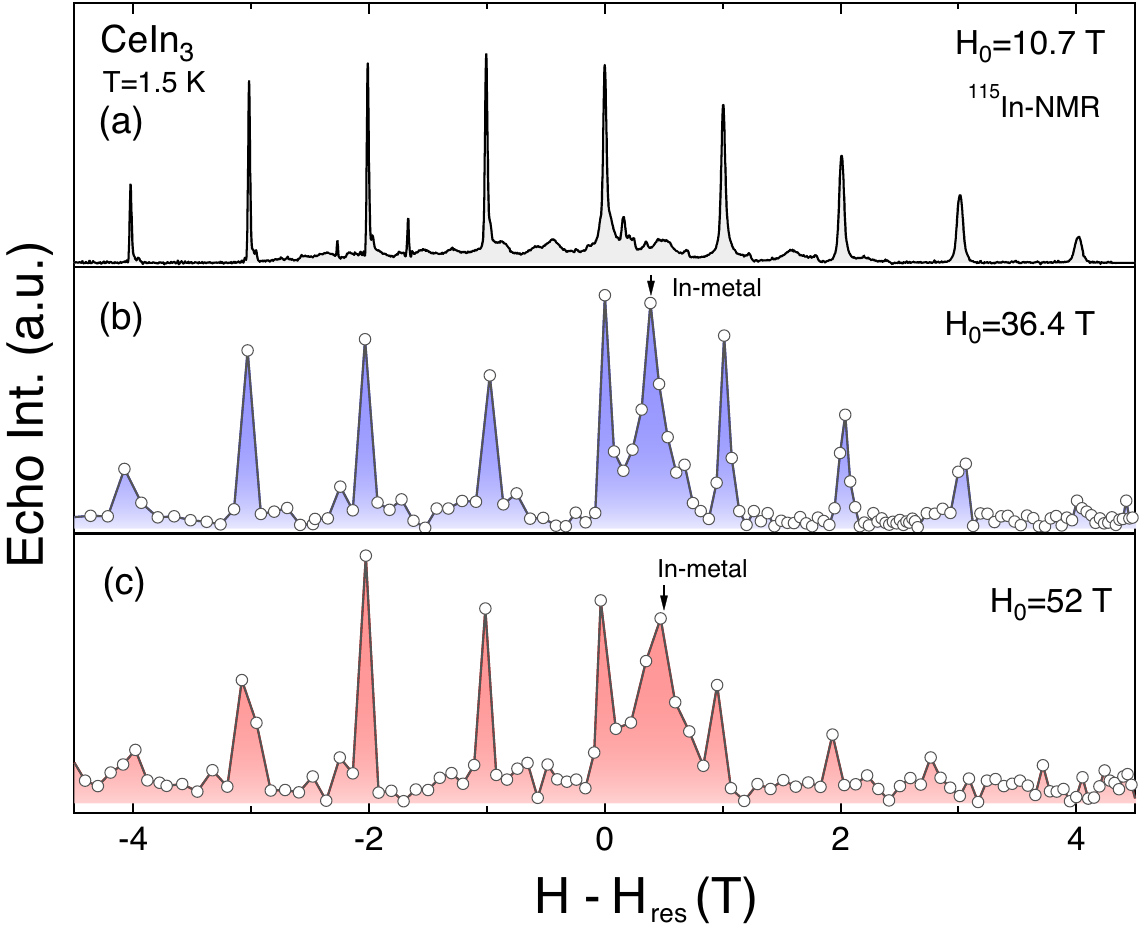}
	\caption{\label{CeIn3_Fig5} Field dependence of the $^{115}$In NMR spectrum of CeIn$_3$ at 1.5 K. Here, $H_{res}$ is the field value of the central peak in each spectrum, i.e., (a) 10.7 T, (b) 36.4 T, and (c) 52 T, respectively. Reproduced with permission from Tokunaga \textit{et al.}, Phys. Rev. B \textbf{99}, 085142 (2019). Copyright 2019 American Physical Society \cite{Tokunaga2019}.}
\end{figure}

CeIn$_3$ is a heavy-fermion compound with a simple cubic crystal structure. It exhibits antiferromagnetic (AFM) order below a N\'{e}el temperature of $T_N = 10$ K. The application of a magnetic field induces a quantum phase transition (QPT) at a critical field $H_c$, which varies between 60 and 80 T, depending on the field orientation. However, the precise nature of this transition remains elusive.
In tunnel-diode-oscillator measurements, an anomaly is detected at $H^{\ast} \sim 45$ T, distinct from results of resistivity measurements. This suggests a possible change in either the magnetic or the crystal structure at $H^{\ast}$. The anomaly could be indicative of a Fermi-surface reconstruction, and has been discussed as the consequence of a Lifshitz transition or a metamagnetic transition, both of which can be driven by a strong magnetic field.
Alternatively, the anomaly at 45 T may signal a transition to a density-wave state, similar to that observed in CeRhIn$_5$.

Tokunaga \textit{et al.} have investigated the high-field phase diagram of CeIn$_3$ using $^{115}$In PFNMR with a spatially homogeneous 60 T pulsed-field magnet (see Ref. \cite{Tokunaga2019} and citations therein). The magnetic field was applied parallel to the crystallographic $\left< 100 \right>$ direction. 
The $^{115}$In nuclei have a spin value of $I = 9/2$, a nuclear gyromagnetic ratio of $\gamma_N/2\pi = 9.3295$ MHz/T, and a large natural abundance of 95.7 $\%$, facilitating a high-sensitivity NMR detection. 

Notably, a high-quality single crystal of CeIn$_3$ becomes highly conductive at low temperatures. Therefore, a challenge encountered in performing PFNMR on such a metallic crystal is the generation of eddy currents induced by the RF pulses, leading to heating effects that make it difficult to probe the intrinsic properties at low temperatures. To minimize the heating, a single crystal of CeIn$_3$ was sliced into thin plates, which were then restacked and separated by thin insulating tapes. 
This methodology provided a pathway for conducting PFNMR on a metallic single crystal at temperatures as low as 1.5 K.

During a magnetic-field pulse, an echo-pulse sequence consisting of 1.5 and 3 $\mu$s RF excitation pulses, separated by 4 $\mu$s, was repeated continually every 200 $\mu$s. 
The full NMR spectrum was then constructed by plotting the integrated intensities of the recorded spin-echo signals against the average value of the time-dependent field during each echo-pulse sequence. For each magnetic-field pulse, the field profile was recorded via the voltage of a pickup coil. The absolute field values were precisely calibrated using $^{63,65}$Cu and $^{115}$In NMR signals from metallic copper and indium powder with known internal fields, located within the same NMR coil as the CeIn$_3$ sample.

\begin{figure}[tbp]
	\centering
	\includegraphics[width=0.9\linewidth]{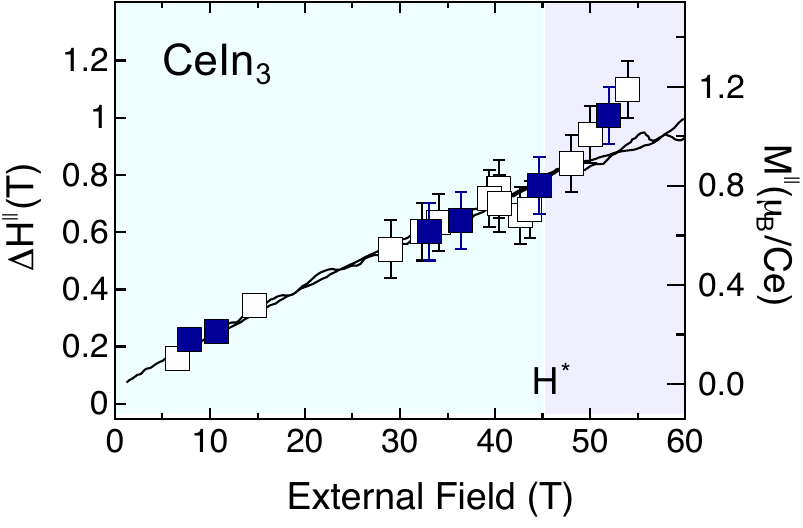}
	\caption{\label{CeIn3_Fig6} Field dependence of the NMR shift $\Delta H^{\parallel} = H_{bare} - H_{res}$ in CeIn$_3$, plotted together with the magnetization $M^{\parallel}$ of a powder sample (solid line, right vertical axis). The values of $H_{res}$ were extracted from the field values of the center (solid squares) and satellite (open squares) peaks in several $^{115}$In NMR spectra measured in different field regions. Reproduced with permission from Tokunaga \textit{et al.}, Phys. Rev. B \textbf{99}, 085142 (2019). Copyright 2019 American Physical Society \cite{Tokunaga2019}.}
\end{figure}

Figure \ref{CeIn3_Fig5} compares $^{115}$In NMR spectra from the AFM ordered state at various fields centered on the resonance field $H_{res}$ of the central peak. For fields of (a) $H_{res} = 10.7$ T and (b) 36.4 T, the spectra are recorded below $H^{\ast}$, whereas (c) 52 T is above $H^{\ast}$. Despite the low resolution in the pulsed-field spectra [(b) and (c)], the spectral structure is preserved at all field regions across $H^{\ast}$, indicating no significant changes in the symmetry or parameters of the crystal lattice. Furthermore, the absence of line broadening or splitting in the NMR spectra above $H^{\ast}$ shows the persistence of hyperfine-field cancellations from Ce-nearest-neighbor moments, indicating that the magnetic structure does not change. There is no sign of incommensurability of the magnetic structure above $H^{\ast}$ either. Although a magnetic-field-induced density-wave scenario is rather unlikely, the resolution of the PFNMR spectra may not be sufficient to conclusively exclude this scenario.

As depicted in Fig.~\ref{CeIn3_Fig6}, the field dependence of the NMR shift $\Delta H^{\parallel} = H_{bare} - H_{res}$, where $H_{bare} = f_{NMR}/^{115}\gamma$, deviates from that of the bulk magnetization $M^{\parallel}$ above $H^{\ast} \simeq 45$ T. This implies a field-induced anomaly of the hyperfine coupling above $H^{\ast}$. Here, the hyperfine coupling is dominated by an indirect mechanism mediated by the orbital hybridizations between Ce and In atoms. These results would be compatible with a field-induced level crossing at $H^{\ast}$.

The PFNMR measurements on metallic single crystals of CeIn$_3$ at low temperatures establish the efficacy of microscopic investigations of heavy-fermion materials at high magnetic fields.

\section{4. Future perspectives}

\subsection{Improvements in PFNMR instrumentation}

PFNMR enables the study of materials by resolving spectral features and nuclear relaxation times as facilitated through the characteristics of pulsed-field magnets. Special coil designs are instrumental in generating spatially homogeneous fields, as was established, for example, at the Laboratoire National des Champs Magnétiques Intenses (LNCMI) in Toulouse \cite{Orlova2016}. An optimal PFNMR magnet for research in condensed-matter physics is characterized by its ability to produce a spatially homogeneous magnetic field with a variation of  $\Delta H / H_0 \approx 10^{-5}$ or better across a volume of several cubic millimeters at the magnet center. Mechanical forces, notably Lorentz forces, induce small deformations during field pulses, requiring magnet designs to account for these deformations to maintain optimal uniformity at peak-field strengths.

Further improvements could involve creating specifically tailored pulsed-field profiles. For example, insert coils with active compensation can generate pulsed-field profiles with constant or controlled linear time-dependent field regimes \cite{Kohama2015}, as discussed in section 2.  
Moreover, employing specially designed capacitors, such as supercapacitors, could be beneficial for realizing long field-pulse durations \cite{Matsui2021}.
Another approach could involve the use of several capacitor units, triggered sequentially to generate multiple maxima of the pulsed field at successive times. This would allow the system to cycle through resonant conditions multiple times during a single magnetic-field pulse.

The peak field of the pulsed magnet should significantly exceed the current 45 T limit of high-field hybrid technology. Furthermore, the magnet coil should be lightweight and designed for rapid cooldown periods between pulses, enhancing experimental efficiency and minimizing the required pulse energy, thereby also increasing safety by reducing the likelihood of critical coil failures.

In terms of the spectrometer's radiofrequency components, capabilities such as high sampling rates, ample analog bandwidth, rapid frequency shifts, and real-time recording capabilities are essential. Using short low-loss cables for all connections and pre-amplifiers close to the resonator is crucial to minimize signal loss at high frequencies and maintain data quality.

Although it should clearly be feasible to perform experiments at temperatures below the currently achieved 1.5 K using pumped $^4$He, the low cooling power of cryostats using $^3$He or a $^3$He/$^4$He mixture presents challenges due to RF heating from the NMR pulses. Moreover, many high-field phenomena may not require the lowest cryogenic temperatures. In some materials with strong internal magnetism, magnetocaloric effects can facilitate sample cooling during the magnetic-field pulse.

Other advancements in PFNMR could be achieved by adopting modern sample-preparation techniques to enhance the instrumental sensitivity. Incorporating microcoils, i.e., NMR coils with an inner diameter of the order of hundred micrometers, would allow for the measurement of microscopic samples. Due to the inherently small spatial field variation across such small samples, this approach would aid in achieving high spectral resolution. 

In order to enhance spectral resolution and to compensate for the deformation of the magnet during field pulses, the implementation of active shimming could be explored. In cases of narrow-line spectra, a truncation in the time-domain spectrum may occur, attributable to the finite duration under resonant conditions. To overcome this limitation and achieve a well-resolved spectrum, advanced data-analysis procedures need to be employed, such as demodulation techniques and sophisticated extrapolations of the time-domain NMR spectra.

The efficacy of alternative NMR detection techniques, such as optically or resistively detected NMR in pulsed magnetic fields, hinges on their unique advantages for specific applications. For instance, optically detected NMR, which incorporates an optical channel in the experimental setup, offers enhanced sensitivity. Resistively detected NMR can be advantageous under extreme conditions and provide valuable insights into materials properties. It also enables NMR measurements of thin-film samples, provided that the hyperfine coupling is sufficiently strong to affect the transport coefficients upon nuclear resonance.

Combining PFNMR with other experimental techniques such as electrical transport, dilatometry, bulk magnetization, or measurements of the magnetocaloric effect within the same sample preparation could, for example, allow for simultaneous investigations of microscopic and macroscopic or thermodynamic quantities in the regime of field-driven phase transitions.

Given the high energies of typically several megajoules involved to produce highest pulsed magnetic fields, it is very likely that high-field PFNMR experiments will remain confined to specialized, large-scale facilities. However, a viable path towards developing a compact, commercially available PFNMR system could be envisioned. Such a setup would provide an alternative to traditional superconducting magnet spectrometers, particularly for analyzing materials characterized by short $T_1$ relaxation times. The proposed concept involves using a miniaturized pulse magnet, capable of generating magnetic fields up to the regime of 30 T. This magnet could be housed within a standard $^4$He cryostat or used as an insert in a wide-bore, high-resolution NMR magnet. Prior to activating the pulse field generated by the insert coil, the steady magnetic field provided by the outer superconducting magnet would be used to pre-polarize the nuclear spin system. Focused on solid-state samples, this approach could facilitate measurements in combined steady and pulsed magnetic fields reaching a total between 40 and 50 tesla. However, the very short total pulse duration of only a few ms of such a miniaturized magnet may affect the spectral resolution through pronounced modulation and truncation effects on the time-domain NMR spectra.

\subsection{Future Research Topics for PFNMR}

High-field states of matter offer a broad range of research topics that can be explored using PFNMR. For example, studying the suppression of superconductivity in high-temperature superconductors at highest magnetic fields is particularly intriguing. This encompasses the study of high-field phase diagrams of unconventional superconductors, which exhibit complex coupling mechanisms and symmetry-breaking phenomena, as reported for compounds such as UTe$_2$ \cite{Aoki2019,Ran2019,Lewin2023}.

Moreover, exotic phenomena of superconductivity at high fields, such as the Fulde-Ferrell-Larkin-Ovchinnikov (FFLO) state \cite{FuldePR1964,LarkinSP1965}, have been reported for several materials, including organic conductors, iron-based superconductors, and other intermetallic compounds. FFLO superconductivity represents a unique state of matter where superconducting pairing occurs with non-zero momentum, resulting in a spatially modulated superconducting order parameter. It has until now primarily been studied using thermodynamic probes and warrants further investigation through PFNMR to provide microscopic information on the spatially modulated electronic spin polarization and the distribution of low-energy quasiparticle densities in the FFLO state.

Materials that exhibit magnetic frustration represent another significant and expansive field of study that can be effectively explored using PFNMR \cite{Lacroix2011}. Magnetic frustration occurs when the magnetic degrees of freedom within a lattice cannot simultaneously minimize their interaction energy, resulting in a degenerate ground state and the absence of conventional long-range magnetic order. 

The materials with magnetic frustration are, for example, of metallic, metal-organic or oxidic composition. The effective dimensionality of the magnetic interactions can range from one-dimensional to three-dimensional.
Of particular fascination are cases where the magnetic interactions can be described using fundamental models of quantum magnetism. This correspondence facilitates empirical validation of fundamental theoretical predictions about unusual ground states and excitations within the collective electronic spin system. Conversely, the experimental discovery of new phenomena can stimulate new theoretical research topics. 

The application of high magnetic fields can stabilize novel magnetic phases, providing a means to differentiate between competing theoretical predictions. By varying the field strength, one can explore phase transitions and quantum critical points, thereby deepening the understanding of the magnetic phase diagrams.
The application of PFNMR to materials with magnetic frustration can uncover detailed information about the local magnetic environments, specifically the distribution of spin polarization within the crystal lattice. Furthermore, $T_1$ relaxation measurements enable the investigation of exotic excitations within these high-field states and the study of the critical slowing down of electronic excitations near field-driven quantum phase transitions. This also applies to materials that exhibit topical phenomena such as spin-ice states or kagome-lattice structures, where frustration leads to exotic ground states and emergent phenomena, including spin liquids.

The Jaccarino-Peter effect is an intriguing physical phenomenon that involves the counterbalancing of an external magnetic field with internal effective exchange fields acting on conduction electrons through exchange interactions with localized magnetic moments \cite{Jaccarino1962}. Typically, these large internal fields suppress superconductivity when no external field is applied. However, in certain magnetic metals, where the internal field opposes the external field, it is possible to neutralize the polarization of conduction electrons. This may allow superconductivity to occur in a region of the phase diagram where these fields compensate each other. By precisely adjusting the external magnetic field to offset the internal magnetism, superconductivity can be re-established. Consequently, high magnetic fields are essential for investigating the Jaccarino-Peter effect.
NMR is established as a powerful technique for probing superconducting states, which may be characterized by spin-singlet coupling and low-energy excitation gaps. PFNMR could be employed to study the Jaccarino-Peter effect by probing alterations in the interplay between the Knight shift, probing the polarization of the conduction electrons, and the polarization of the localized moments with applied magnetic fields. Ideally, these experiments would be complemented by measurements of the electronic-transport properties across the transition to the high-field superconducting state.

The Kondo effect is a quantum phenomenon observed in metals, characterized by the scattering of conduction electrons with localized magnetic moments \cite{Kondo1964}. This eventually leads to the formation of a screened singlet state below a critical Kondo temperature. Hybridization between the itinerant and localized electrons can induce the fomation of a band gap, typically on the order of tens of meV. In some cases, this phenomenon may result in a temperature-dependent transition from metallic to insulating behavior, as evidenced by resistivity measurements. Materials exhibiting this effect, such as SmB$_6$, Ce$_3$Bi$_4$Pt$_3$, or YbB$_{12}$, are classified as Kondo insulators.
High magnetic fields can induce insulator-to-metal transitions in these materials. Employing PFNMR could facilitate the examination of changes in electron density and mobility during these transitions. This includes investigating the impact of electron-electron and electron-phonon interactions, as well as associated critical fluctuations. Specifically, PFNMR may allow for the study of low-energy electronic density of states and the magnetic properties of localized moments by observing variations in the Knight shift and the spin-lattice relaxation time $T_1$ under different magnetic field strengths.

Lastly, continuing from the discussion in section 3, PFNMR could be of great benefit for the development of contrast agents for use in high-field MRI scanners. These applications require contrast agents with elevated relaxivity at high NMR frequencies to ensure distinct and clear imaging results. The implementation of flat-top PFNMR measurements of $T_1$ relaxation times could significantly improve the design and synthesis of next-generation contrast agents. Consequently, contrast agents optimized for high magnetic fields could be developed early, aligning with advances in magnet technology that will enable the use of MRI scanners with significantly higher steady magnetic fields. By tailoring of contrast agents to optimize their performance at higher magnetic fields, one could achieve more precise imaging, aiding in the accurate diagnosis and assessment of a wide range of medical conditions. Furthermore, investigating possible interactions between contrast agents and biological tissues is an important topic. This interaction analysis is essential for ensuring the safety and efficacy of contrast agents, particularly in high magnetic fields, which may alter molecular dynamics and properties. Advancing PFNMR techniques in line with emerging needs in high-field MRI technology not only enhances the functionality and application range of MRI contrast agents, but also contributes to the broader field of medical imaging by enabling more detailed and accurate studies of physiological and pathological processes.

\section{5. Conclusions}
This article introduces and highlights the role of nuclear magnetic resonance in strong pulsed magnetic fields in advancing research within solid-state physics, materials science, and high-resolution NMR. Originating from developments in several laboratories over recent years, PFNMR has matured and is now instrumental in probing the complexities of materials exposed to extreme magnetic fields. It facilitates the study of materials with intricate interactions and strong electronic correlations, particularly when subjected to fields beyond the capacity of steady-field magnets. PFNMR has already provided novel insights into phenomena such as spin-nematic phases, magnetic superlattices, and field-induced changes of electronic properties in metals through the investigation of materials such as LiCuVO$_4$, SrCu$_2$(BO$_3$)$_2$, and CeIn$_3$. 

With its broad applicability extending to high-$T_c$ superconductors, materials with strong exchange coupling, and systems undergoing high-field transitions, PFNMR promises to continue uncovering new states of matter and deepening our understanding of quantum materials. Anticipated instrumental advancements are expected to further widen  the utility of PFNMR, making it increasingly available to the global scientific community.

\section*{Acknowledgments}
We extend our gratitude to several colleagues for their invaluable support in the preparation of this manuscript. Their numerous comments, suggestions for improvement, and permissions to use figures have been instrumental. In particular, we acknowledge Tomoki Kanda, Sven Luther, Florian B\"{a}rtl, Tommy Kotte, Yo Tokunaga, J\"{u}rgen Haase, Anna Orlova, Geert Rikken, Raivo Stern, Jochen Wosnitza, and Yoshimitsu Kohama.
We acknowledge the support of HLD at HZDR, member of the European Magnetic Field Laboratory (EMFL).
This work is partly supported by the JSPS KAKENHI under Grant Nos. 19H01832, 22H00104, 22H04458, and 24H01599. 

\section*{Notes on contributors}
\begin{wrapfigure}{l}{0.1\textwidth}
\vspace{-10pt}
\includegraphics[width=0.11\textwidth]{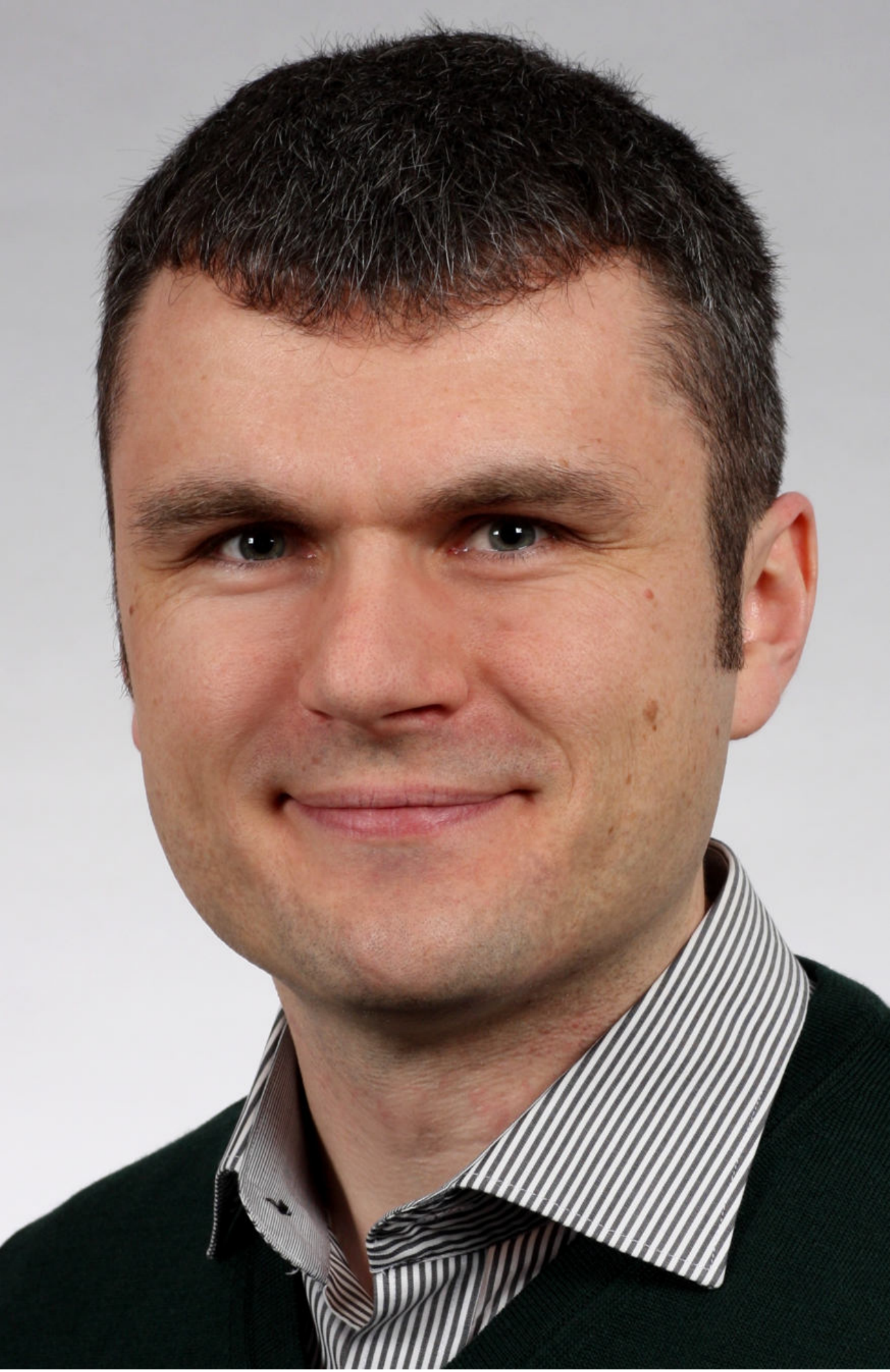}
\end{wrapfigure}
Hannes K\"{u}hne obtained his PhD in physics from the Dresden University of Technology in 2011. After a postdoctoral position at the National High Magnetic Field Laboratory in Tallahassee, he moved to the Dresden High Magnetic Field Laboratory in 2013 as a specialist in NMR of condensed-matter systems under extreme environmental conditions. His research interests include the investigation of phenomena in unconventional superconductivity and low-dimensional quantum spin systems.
\\

\begin{wrapfigure}{l}{0.1\textwidth}
\vspace{-10pt}
\includegraphics[width=0.11\textwidth]{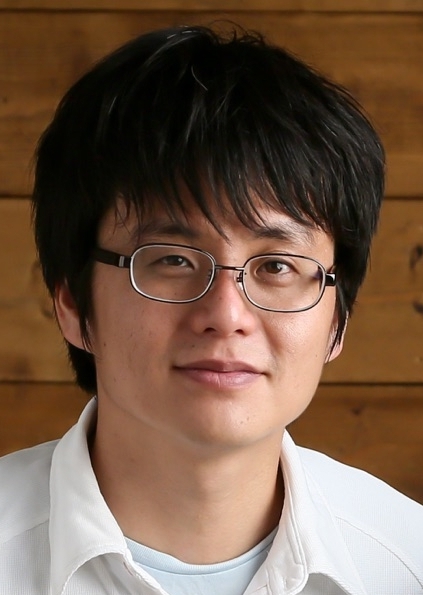}
\end{wrapfigure}
Yoshihiko Ihara obtained his PhD in science from the Kyoto University in 2008. After a postdoctoral position at the Laboratoire de Physique des Solides in Paris-Sacley, he moved to the Hokkaido University in 2010. His research interests include the emergent quantum phenomena in strongly correlated electron systems and technical development to capture the exotic properties of electrons in extreme conditions.


\begin{thebibliography}{99}

\bibitem{Bloch1946}
F. Bloch, W. W. Hansen, and M. Packard, Phys. Rev. \textbf{69}, 127 (1946).

\bibitem{Purcell1946}
E. M. Purcell, H. C. Torrey, and R. V. Pound, Phys. Rev. \textbf{69}, 37 (1946).

\bibitem{Abragam1961}
A. Abragam, \textit{The Principles of Nuclear Magnetism} (Oxford University Press, London, 1961).

\bibitem{Fukushima1981}
E. Fukushima and S. B. W. Roeder, \textit{Experimental Pulse NMR - A Nuts and Bolts Approach} (Westview Press, 1981).

\bibitem{Mehring1983}
M. Mehring, \textit{Principles of High Resolution NMR in Solids} (Springer Verlag, 1983).

\bibitem{Slichter1990}
C. P. Slichter, \textit{Principles of Magnetic Resonance} (Springer Verlag, 1990).

\bibitem{Matsui2021}
K. Matsui, T. Kanda, Y. Ihara, K. Kindo, and Y. Kohama, Rev. Sci. Instrum. \textbf{92}, 024711 (2021).

\bibitem{Miura2003}
N. Miura, T. Osada, and S. Takeyama, J. Low Temp. Phys. \textbf{133}, 139 (2003).

\bibitem{Haase2003}
J. Haase, D. Eckert, H. Siegel, H. Eschrig, K. H. M\"{u}ller, and F. Steglich, Solid State Nucl. Magn. Reson. \textbf{23}, 263 (2003).

\bibitem{Haase2004}
J. Haase, Appl. Magn. Reson. \textbf{27}, 297 (2004).

\bibitem{Kozlov2005}
M. B. Kozlov, J. Haase, C. Baumann, and A. G. Webb, Solid State Nucl. Magn. Reson. \textbf{28}, 64 (2005).

\bibitem{Haase2005}
J. Haase, M. B. Kozlov, A. G. Webb, B. B\"{u}chner, H. Eschrig, K.-H. M\"{u}ller, and H. Siegel, Solid State Nucl. Magn. Reson. \textbf{27}, 206 (2005).

\bibitem{Zheng2009}
G.-Q. Zheng, K. Katayama, M. Ishiyama, S. Kawasaki, N.
Nishihagi, S. Kimura, M. Hagiwara, and K. Kindo, J. Phys. Soc.
Jpn. \textbf{78}, 095001 (2009).

\bibitem{Hamad2011}
E. Abou-Hamad, P. Bontemps, and G. L. J. A. Rikken, Solid
State Nucl. Magn. Reson. \textbf{40}, 42 (2011).

\bibitem{Meier2011}
B. Meier, S. Greiser, J. Haase, T. Herrmannsd\"{o}rfer, F. Wolff-Fabris, and J. Wosnitza, J. Magn. Reson. \textbf{210}, 1 (2011).

\bibitem{Meier2012}
B. Meier, J. Kohlrautz, J. Haase, M. Braun, F. Wolff-Fabris, E. Kampert, T. Herrmannsd\"{o}rfer, and J. Wosnitza, Rev. Sci. Instrum. \textbf{83}, 083113 (2012).

\bibitem{Stork2013}
H. Stork, P. Bontemps, and G. L. J. A. Rikken, J. Magn. Reson.
\textbf{234}, 30 (2013).

\bibitem{Kohlrautz2016}
J. Kohlrautz, S. Reichardt, E. L. Green, H. K\"{u}hne, J. Wosnitza,
and J. Haase, J. Magn. Reson. \textbf{263}, 1 (2016).

\bibitem{Orlova2017}
A. Orlova, E. L. Green, J. M. Law, D. I. Gorbunov, G. Chanda, S. Kr\"{a}mer, M. Horvati\'c, R. K. Kremer, J. Wosnitza, and G. L. J. A. Rikken, Phys. Rev. Lett. \textbf{118}, 247201 (2017).

\bibitem{Kohlrautz2016SCBO}
J. Kohlrautz, J. Haase, E. L. Green, Z. T. Zhang, J. Wosnitza,
T. Herrmannsd\"{o}rfer, H. A. Dabkowska, B. D. Gaulin, R. Stern,
and H. K\"{u}hne, J. Magn. Reson. \textbf{271}, 52 (2016).

\bibitem{Tokunaga2019}
Y. Tokunaga, A. Orlova, N. Bruyant, D. Aoki, H. Mayaffre, S. Kr\"{a}mer, M.-H. Julien, C. Berthier, M. Horvati\'c, N. Higa, T. Hattori, H. Sakai, S. Kambe, and I. Sheikin, Phys. Rev. B \textbf{99}, 085142 (2019).

\bibitem{Orlova2016}
A. Orlova, P. Frings, M. Suleiman, and G. L. J. A. Rikken, J.
Magn. Reson. \textbf{268}, 82 (2016).

\bibitem{Ihara2021}
Y. Ihara, K. Hayashi, T. Kanda, K. Matsui, K. Kindo, and Y. Kohama, Rev. Sci. Instrum. \textbf{92}, 114709 (2021).

\bibitem{Kohama2022}
Y. Kohama, T. Nomura, S. Zherlitsyn, and Y. Ihara, J. Appl. Phys. \textbf{132}, 070903 (2022).

\bibitem{Wei2023}
W. Wei, Q. Liu, L. Yuan, J. Zhang, S. Liu, R. Zhou,
Y. Luo, and X. Han, IEEE Trans. Instrum. Meas.,
\textbf{72}, 1 (2023).

\bibitem{Kohama2015}
Y. Kohama, and K. Kindo, Rev. Sci. Instrum. \textbf{86}, 104701 (2015).

\bibitem{Bednorz1986}
J. G. Bednorz and K. A. M\"{u}ller, Z. Phys. B \textbf{64}, 189 (1986).

\bibitem{Livas2020}
J. A. Flores-Livas, L. Boeri, A. Sanna, G. Profeta, R. Arita, and M. Eremets, Phys. Rep. \textbf{856}, 1 (2020). 

\bibitem{Schollwoeck2004}
U. Schollw\"{o}ck, J. Richter, D. J. J. Farnell, and R. F. Bishop, eds., \textit{Quantum magnetism}
(Springer, Berlin, 2004).

\bibitem{Din2023}
R. N. Din, \textit{Nuclear magnetic resonance studies of paramagnetic relaxation enhancement at high magnetic fields: methods and applications}, PhD thesis, Université Grenoble Alpes, 2023.


\bibitem{Aoki2019}
D. Aoki \textit{et al.}, J. Phys. Soc. Jpn. \textbf{88}, 043702 (2019).

\bibitem{Ran2019}
S. Ran \textit{et al.}, Science \textbf{365}, 684 (2019).

\bibitem{Lewin2023}
S. K. Lewin, C. E. Frank, S. Ran, J. Paglione, and N. Butch, Rep. Prog. Phys. \textbf{86} 114501 (2023).

\bibitem{FuldePR1964}
P. Fulde and R. A. Ferrell, Phys. Rev. {\bf 135}, A550 (1964).

\bibitem{LarkinSP1965}
A. I. Larkin and Y. N. Ovchinnikov, Sov. Phys. JETP {\bf 20}, 762 (1965).

\bibitem{Lacroix2011}
C. Lacroix, P. Mendels, and F. Mila, \textit{Introduction to Frustrated Magnetism}, Springer Series in Solid-State Sciences, volume 164, 2011.

\bibitem{Jaccarino1962}
V. Jaccarino and M. Peter, Phys. Rev. Lett. \textbf{9}, 290 (1962).

\bibitem{Kondo1964}
J. Kondo, Prog. Theor. Phys. \textbf{32}, 37 (1964).



\end{thebibliography}
\end{document}